\documentclass[lettersize,journal]{IEEEtran}
\usepackage{amsmath,amsfonts}
\usepackage{algorithmic}
\usepackage{algorithm}
\usepackage{array}
\usepackage[caption=false,font=normalsize,labelfont=sf,textfont=sf]{subfig}
\usepackage{textcomp}
\usepackage{stfloats}
\usepackage{url}
\usepackage{verbatim}
\usepackage{graphicx}
\usepackage{cite}
  \usepackage{epstopdf}
\usepackage{caption}
\usepackage{color}
\usepackage{bm}

\begin{document}

\title{Collaborative Authentication for 6G Networks:  An Edge Intelligence based Autonomous  Approach}

\author{He~Fang,~\IEEEmembership{Member, IEEE},~Zhenlong~Xiao,~\IEEEmembership{Member, IEEE},~Xianbin~Wang,~\IEEEmembership{Fellow, IEEE}, Li~Xu,~\IEEEmembership{Member, IEEE}, and~Lajos~Hanzo,~\IEEEmembership{Life Fellow, IEEE}

\thanks{This work was supported in part by the National Natural Science Foundation of China under Grants 62271430; in part by the Natural Sciences and Engineering Research Council of Canada (NSERC) Discovery Program under Grant RGPIN2018-06254, and  the Canada Research Chair Program; in part by  the Engineering and Physical Sciences Research Council projects EP/W016605/1 and EP/X01228X/1, and the European Research Council's Advanced
Fellow Grant QuantCom (Grant No. 789028).}
\thanks{H. Fang is with the School of Electronic and Information Engineering, Soochow University, Soochow 215301, China  (email: fanghe@suda.edu.cn)}
\thanks{Z. Xiao is with the Department of Informatics and Communication
Engineering, School of Informatics, Xiamen University, Xiamen 361005, China (email: zlxiao@xmu.edu.cn)}
\thanks{X. Wang is with the Department of Electrical and Computer Engineering, Western University, London, ON N6A 5B9, Canada (email: xianbin.wang@uwo.ca)}
\thanks{L. Xu is with the College of Computer and Cyber Security and the Fujian Provincial Key Laboratory of Network Security and Cryptology, Fujian Normal University, Fuzhou 350117, China (email: xuli@fjnu.edu.cn)}
\thanks{L. Hanzo is with School of Electronics and Computer Science, University of Southampton, SO17 1BJ, U.K (email: lh@ecs.soton.ac.uk)}

}

\maketitle

\begin{abstract}
The conventional device authentication of wireless networks usually
relies on a security server and centralized process, leading to long
latency and risk of single-point of failure.  While these challenges
might be mitigated by collaborative authentication schemes, their
performance remains limited by the rigidity of data collection and
aggregated result.  They also tend to ignore attacker localization in
the collaborative authentication process. To overcome these
challenges, a novel collaborative authentication scheme is proposed,
where multiple edge devices act as cooperative peers to assist the
service provider in distributively authenticating its users by estimating their received signal strength
indicator (RSSI) and mobility trajectory
(TRA).  More explicitly, a distributed
  learning-based collaborative authentication algorithm is conceived,
  where the cooperative peers update their authentication models
  locally, thus the network congestion and response time remain low.  Moreover, a
situation-aware secure group update algorithm is proposed for
autonomously refreshing the set of cooperative peers in the dynamic
environment.  We also develop an
  algorithm for localizing a malicious user by the cooperative peers once it
  is identified. The simulation results demonstrate that
the proposed scheme is eminently suitable for both indoor and outdoor
communication scenarios, and outperforms some existing benchmark
schemes.
\end{abstract}


\IEEEpeerreviewmaketitle

\section{INTRODUCTION}
The sixth generation (6G) technologies are expected to evolve from
personal communication towards the full realization of the Internet of
Things (IoT) paradigm, interconnecting not only people, but also
machines, vehicles, computing resources, industry processes, and even
robotic agents \cite{08}.  Due to the open broadcast nature of radio
signal propagation and the standardized transmission schemes used,
wireless communications are extremely vulnerable to security
threats~\cite{81,07,01}. More specifically, the highly heterogeneous
network structure, time-varying topology, and ubiquitous
resource-constraint devices used in 6G networks leave many loopholes
for potential eavesdropping, spoofing, forgery, interception, and
denial of service attacks \cite{36,37,41}. These security threats may
lead to privacy leakage, interruption of intelligent services, and
even to overall system breakdown. In preventing these
  malicious attacks, authentication is imperative to confirm the
  identities of communicating devices, to check the validity of their
  access to the network, to maintain the integrity and trustworthiness of their
  communications~\cite{35}.

\subsection{Challenges for Existing Authentication Methods}
The conventional authentication methods usually
  apply classic symmetric/asymmetric-key
  cryptography~\cite{09,70,71,72}.  The Public Key Infrastructure
(PKI) has been widely studied in the literature \cite{03,16}, where
the security relies on a set of globally trusted authorities.  The
identity-based signature techniques allow a user's public key to be
readily computed from its known identity information, thus eliminating
the need for public-key/certificates \cite{23}.  The Diffie-Hellman
key agreement protocol is known to be vulnerable to the
``man-in-the-middle" attack, if two users involved in the protocol do
not share any authenticated information about each other, e.g. shared keys, certificates, and passwords, prior to the
protocol's execution \cite{12}. The
  information-theoretic secret key agreement between a pair of
  legitimate parties guarantees a certain level of security, while
  relying on no computational restrictions concerning the
  eavesdropper~\cite{11,73,74,75,110,111}. As a design alternative, physical
layer authentication exploits the unique random nature of
communication links and devices-related
features between a pair of transceivers to identify transmitters~\cite{10,24,113}.

\begin{figure*}[htbp]
\centering
\caption*{TABLE I: Contrasting the novelty of the proposed solution to the state-of-the-art.}
\includegraphics[width=18cm,height=3.8cm]{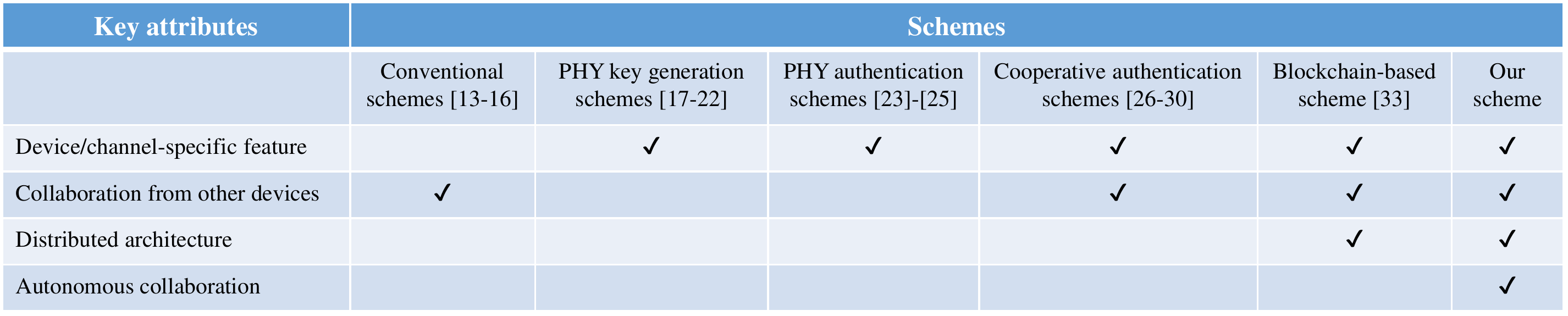}
\end{figure*}

However, most of the existing security mechanisms feature
network-specific, stand-alone, and isolated designs, which are
typically deployed in a particular network, application, and certain
layer of the protocol stack. Such mechanisms typically involve two
parties, where one of the entities has to be authenticated, while the
other one performs the verification. These isolated authentication
schemes validate the devices without any cooperation from other
devices. Their performance usually suffers from the limited knowledge
and computational resources of the specific device performing the
authentication. Moreover, the performance of physical layer authentication
  suffers from low reliability and robustness in highly dynamic
  networks due to the noisy time-varying observations of the radio
  frequency fingerprint.

To overcome the challenges of isolated methods, collaborative security
frameworks have been proposed for solving many practical security
problems  \cite{13,02,14,25,112}. The collaborative methods could provide more
  accurate  and robust authentication decisions by enabling
multi-dimensional information sharing among the collaborative devices. Moreover, security enhancement can be achieved
by increasing the difficulty of being successfully cracked.  The
authors of \cite{02} proposed a cooperative authentication scheme for
underwater acoustic sensor networks, which relies on trusted nodes
that independently help a sink node to evaluate the belief concerning
each incoming packet and then to reach an authentication decision.
A cooperative authentication scheme is designed for
  detecting the spoofed Global Positioning System (GPS) signals
  in~\cite{14}. The authors of~\cite{25} proposed a physical layer
authentication scheme based on the cooperation of multiple landmarks,
which collect the channel information of the transmitter for
authentication.

However, the above centralized methods do not
  consider the risk of single-point failure. Their further limitation
  is that they may suffer from long authentication latency and response
  time when numerous devices request authentication services at the
  same time. To address these issues, developing a distributed method
  to move security management from the cell-center to the cell-edge in
  the network is an efficient alternative. It will provide secure and
  low-overhead authentication by enabling a server to delegate its
  authentication authority to edge nodes~\cite{06,29}.  A blockchain
empowered group authentication scheme is proposed in \cite{26} for
vehicles with decentralized identification based on a secret sharing
and dynamic proxy mechanism.  However, the blockchain-based schemes may
suffer from long latency, high computation and communication cost, as
well as high storage resources required for running a blockchain
\cite{30,31}.

To elaborate a little further on the challenges, the data collection
and aggregation models of typical collaborative security schemes are
usually fixed and stationary. They have limited capability in processing
heterogeneous data, leading to potential failure in capturing the
critical aspects of practical communication environments. These will
also lead to limitations in intelligently exploiting heterogeneous
security information in 6G communications requiring situation-aware
services and flexible processes. More importantly,  the
information uncertainties and complex network topology will impose new
challenges on the reliable authentication in mobile networks, such as
vehicular ad hoc networks (VANETs) and Unmanned Aerial Vehicle (UAV)
networks. In a nutshell, an autonomous distributed
  authentication technique is extremely helpful, where multiple
  devices govern the authentication process by sharing
  multi-dimensional information among them to improve the
  authentication accuracy and reliability.

\subsection{Contributions}
This paper proposes an edge intelligence-based collaborative
authentication scheme for accurate identification and attacker
localization by harnessing the cooperation of multiple edge nodes
(peers). Each cooperative peer helps the service
  provider to authenticate its users by collecting and
  processing their observations of the users'
  location-related features, including the RSSI  and TRA.
The cooperative peers share their local observations with others and
make real-time authentication decision based on a distributed learning
algorithm.  Consensus is achieved in the proposed scheme, where a
group of cooperative peers will reach an agreement in the
  information collection and collaborative authentication
process.  Moreover, the learning models are updated locally at
cooperative peers and  only the final decision will be sent to the
service provider. Then, a situation-aware secure group update
algorithm is developed for adaptively updating the collaborative peers
and the authentication features in the dynamic network. Finally, an
attacker localization algorithm is developed for finding the position
of the user, once it is identified as a spoofer.

The contributions of this paper are summarized as
follows:\\
1) We propose an edge intelligence-based collaborative authentication scheme using multiple cooperating edge nodes to collect location-related features of the user for
  formulating their final decision. The proposed scheme decreases the time latency and network load by moving the security provision from center to edge of the  network. It also enhances security by utilizing multiple cooperating
 devices and by using multi-dimensional security information. \\
2) The situation-aware group update algorithm autonomously updates the cooperative peers and authentication features by adaptively identifying the dynamic environment and network topology.  Hence, the proposed scheme provides high robustness of collaborative authentication in the mobile network. Moreover, the developed attacker localization algorithm can be used to immediately locate an identity spoofer,
thus providing security enhancement. \\
3) Simulation results are provided for both indoor and outdoor communication scenarios. The results show that using more  cooperative devices increases the authentication accuracy at the
  cost of higher communication overhead.  The automation
  in updating the cooperating edge nodes and the robustness of the  proposed scheme are verified in realistic noisy time-varying
  environments. It is also shown that the proposed scheme performs better than some existing counterparts.

\begin{table}[htbp]
\caption*{TABLE II: Symbol definitions of this paper} \label{fig:predicate}
\centering
{\renewcommand\arraystretch{1.5}
\begin{tabular}{|l|l|}
\hline
Symbol   &       Definition \\\hline
$N$   &     Number of cooperative edge nodes (peers). \\\hline
$\bm{H}_{n}$   &     Feature estimation of the user observed by the $n$-th  peer. \\\hline
$I$   &     Identity (ID) of the user to be authenticated. \\\hline
$I_{n}$   &   ID of the user observed by the $n$-th   cooperative peer. \\\hline
$\Phi_{0}$   &     Case that the user to be authenticated is legitimate. \\\hline
$\Phi_{1}$   &     Case that the user to be authenticated is an identity spoofer. \\\hline
$\nu$   &    Collaborative  authentication  threshold. \\\hline
$F_{\rm{MD}}$   &    Misdetection (MD) rate of collaborative authentication. \\\hline
$F_{\rm{FA}}$   &     False alarm (FA) rate of collaborative authentication. \\\hline
$f_{n}(\bm{x}_{n})$   &     Local objective function of the $n$-th  cooperative peer. \\\hline
$\bm{x}_{n}$   &    Local variable of the $n$-th cooperative peer. \\\hline
$\bm{x}_{0}$   &     Agreement of all cooperative peers. \\\hline
$\bm{a}$   &     Real position of the legitimate  user. \\\hline
$\bm{b}_{n}$   &     Real position of the $n$-th cooperative peer. \\\hline
$\bm{c}$   &     Real position of the spoofer. \\\hline
\end{tabular}\label{Predicates}
}
\end{table}

The rest of this paper is organized as follows. In Section II, the system model used  is presented. In Section III, we present the proposed  edge intelligence-based collaborative authentication  scheme. The authentication performance analysis is also  given  in Section III. The simulation results are discussed in Section IV. Finally, Section V concludes the paper. The symbol definitions of this paper are given in TABLE II.

\section{SYSTEM MODEL}

In this paper, we consider a wireless communication system associated with location-based services, where the service provider monitors  its user's  geographical location by positioning techniques, e.g. by the widely used GPS. The system contains massive edge nodes,  e.g.  gateways,  servers, access points, and full-function devices.

\begin{figure}[htbp]
\centering
\includegraphics[width=7cm,height=8cm]{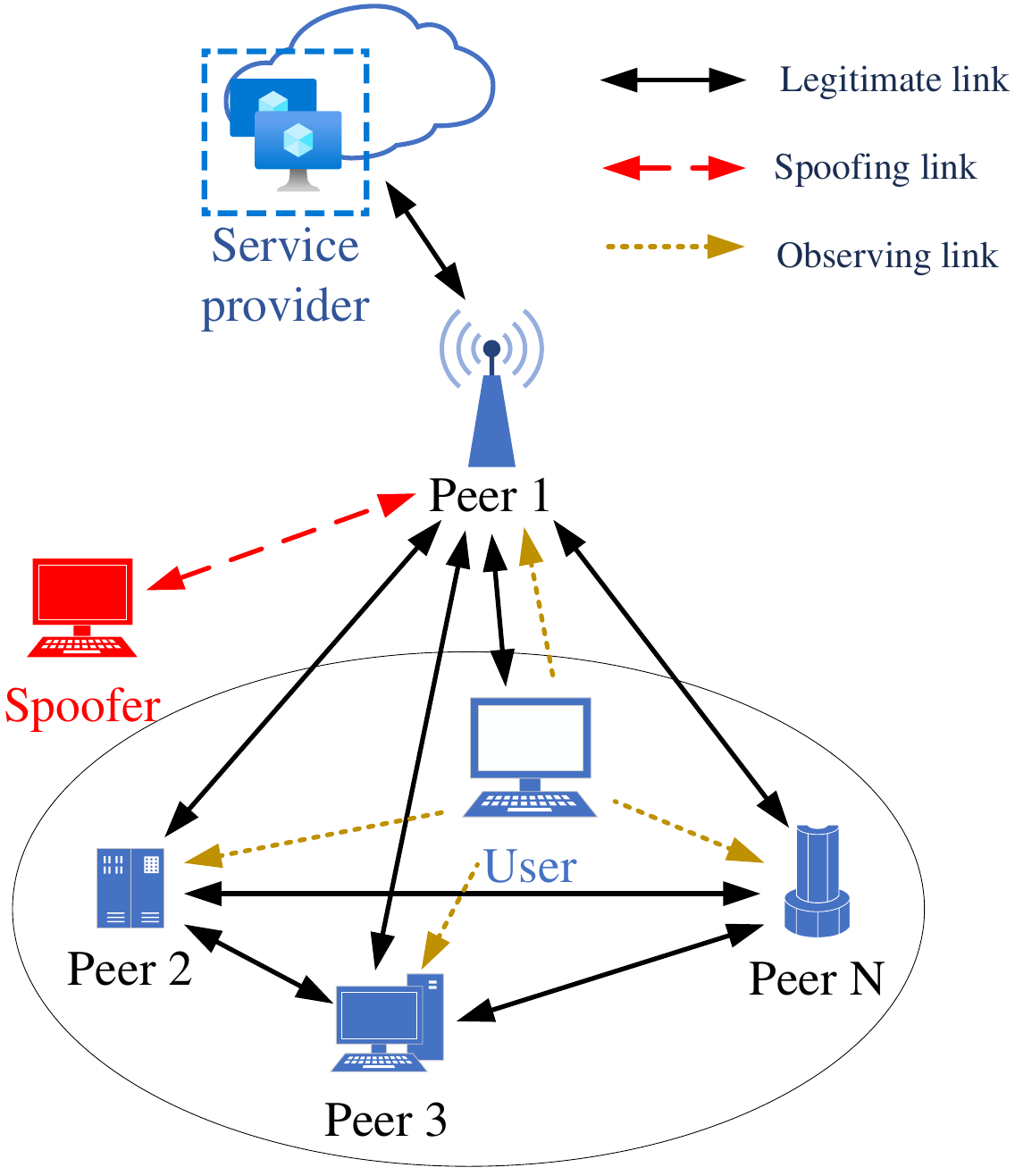}
\caption{A system providing location-based services,
    which suffers from either identity or location spoofing attacks. The
    service provider authenticates its users relying on the
    collaboration of multiple edge nodes near to the user.}
\label{fig:fig13}
\end{figure}

\subsection{Attack Model}

As shown in Fig. 1, the system suffers from security threats caused by  \\
\emph{$\circ$ Identity spoofer:} It performs identity spoofing attacks by  imitating the legitimate user and then seeks to glean  illegal benefits from the service provider;\\
\emph{$\circ$ Location spoofer:} It performs location spoofing attacks by misleading  the service provider through pretending to be at a hypothetical  position, e.g. GPS spoofer \cite{14}.

The attack  model considered in this paper is similar to the traditional  Alice-Bob-Eve model \cite{07,01,36}, the service provider can be seen as Bob who needs to identify its users, while a user can be seen as Alice or Eve.
The following assumptions are stipulated concerning  the attack model:  \\
\textit{Assumption 1.} The attacker is a single malicious node, capable of  performing any kind of signal processing techniques, and it has unlimited transmission power
capabilities;\\
\textit{Assumption 2.} The attacker knows the positions of all  edge nodes and of the legitimate user;\\
\textit{Assumption 3.} The attacker is located at a different position  from the legitimate user, which is a smart device and focuses on maximizing its utility.

Note that when the spoofer is located close to
  the legitimate user, it may be readily spotted by the user. In order
  to maximize its utility gained, the spoofer will be located at an
  appropriate distance from the user.  Hence, the
  above assumptions are reasonable. Moreover, guarding against
  a powerful worst-case attacker clearly demonstrates the benefits of
  our scheme.
The service provider has to authenticate its user to confirm that it is a legitimate device and it is indeed located at  the reported position.
The authentication between the service provider and its user  is  assisted by $N$ trusted but heterogeneous edge nodes, which  are located near to the user at the network edge.
These edge nodes can be termed  as cooperative peers, which  collect information of the user to be authenticated and help the service provider to verify  its user's identity and localization.

\subsection{Location-related Features for Collaborative Authentication}
Two location-related features are utilized for collaborative
authentication in this paper, i.e. the RSSI and TRA. Specifically, the RSSI is determined by the transmission
power, the distance between the transceivers, and the radio
environment \cite{01}. Path loss models \cite{21}  are typically of the form
\begin{eqnarray}
PL=A\log_{10}(D)+B+C\log_{10}(\kappa_{c}/5),
\end{eqnarray}
where $D$ is the distance between the transceivers, $\kappa_{c}$ is the carrier  frequency, the fitting parameter $A$ includes the path-loss exponent,
$B$ is the intercept and  environment-specific term, and $C$ describes the path loss frequency dependence.
The TRA  of the user can be observed by   cameras or laser radars \cite{22,27}, which are widely used in the mobile networks, such as VANETs, flying ad-hoc networks (FANETs), aeronautical ad hoc networks (AANETs), and UAV networks.

Each cooperative peer locally estimates one of these features of the   user  for identification. The observation of the $n$-th cooperative peer is denoted as
\begin{eqnarray}
(I_{n},\bm{H}_{n}),~ n\in\{1,2,...,N\},
\end{eqnarray}
where $I_{n}$ is the ID of the user to be authenticated observed by the $n$-th cooperative peer.
\begin{eqnarray}
\bm{H}_{n}=(H_{n1},H_{n2})^{\rm{\dagger}}
\end{eqnarray}
represents the feature estimation of the user to be authenticated at the $n$-th cooperative peer. $\dagger$ denotes the transposition symbol.
If the $i$-th feature is not selected by the $n$-th cooperative peer,  we denote  $H_{ni}=0$, where $i\in \{1,2\}$.  The RSSI  and TRA are listed as the 1st and 2nd feature, respectively.

Given the estimates of the above features, the
  distance  between the user to be authenticated and each
  cooperative peer can be derived. The location of the user to be
  authenticated can be derived based on the estimates of
  location-related features by multiple cooperative peers  \cite{76,77,78}. Note that more location-related features can be selected in different application scenarios, e.g.
angle-of-arrival (AoA). The RSSI and TRA are utilized as examples in this paper.

\subsection{The Proposed Collaborative Authentication System}
Our objective is to achieve accurate authentication and to locate the
attacker based on the cooperation of multiple edge nodes.  Upon
denoting the real position of the legitimate user as
$\bm{a}=(a_{1},a_{2},a_{3})^{{\rm{\dagger}}}$, the collaborative
authentication based on the observations $(I_{n},\bm{H}_{n}),
n=1,2,...,N$, is formulated as
\begin{eqnarray}
\begin{cases}
\forall n, I_{n}=I~{\rm{and}}\parallel\bm{x}_{0}(\bm{H}_{1},...,\bm{H}_{N})-\bm{a}\parallel_{2}\leq\nu & \Phi_{0}\\
\exists n, I_{n}\neq I~{\rm{or}}\parallel\bm{x}_{0}(\bm{H}_{1},...,\bm{H}_{N})-\bm{a}\parallel_{2}>\nu &\Phi_{1}
\end{cases},
\end{eqnarray}
where $\nu$ is a collaborative authentication
  threshold. $\Phi_{0}$ represents that the user to be
authenticated is legitimate, while $\Phi_{1}$ indicates that it is an
identity spoofer. Moreover,
  $\bm{x}_{0}(\bm{H}_{1},\bm{H}_{2},...,\bm{H}_{N})=(x_{01},x_{02},x_{03})^{\rm{\dagger}}$
  represents an agreement of all cooperative peers concerning the
  authentication decision based on the observations of (2). If the
  user to be authenticated cannot be observed by the cooperative
  peers, $\bm{x}_{0}$ is set to
  $(+\infty,+\infty,+\infty)^{\rm{\dagger}}$.
The performance of the proposed scheme is evaluated by the following criteria:
\subsubsection{Authentication Accuracy}
It  can be evaluated by the MD rate and FA rate, which are  formulated, respectively, as
\begin{eqnarray}
F_{\rm{MD}}={\rm{Pr}}(\parallel\bm{x}_{0}(\bm{H}_{1},\bm{H}_{2},...,\bm{H}_{N})-\bm{a}\parallel_{2}\leq\nu\mid \Phi_{1})
\end{eqnarray}
and
\begin{eqnarray}
F_{\rm{FA}}={\rm{Pr}}(\parallel\bm{x}_{0}(\bm{H}_{1},\bm{H}_{2},...,\bm{H}_{N})-\bm{a}\parallel_{2}>\nu\mid \Phi_{0}),
\end{eqnarray}
where ${\rm{Pr}}(\cdot)$ is the
  probability notation. The prerequisite of the above formulation is that the
  user's ID observed by all cooperative peers is correct,
  i.e. $\forall n, I_{n}=I$. Otherwise, the user to be authenticated
  will be directly identified as a spoofer by the cooperative peers,
  i.e. $\exists n, I_{n}\neq I$.

\subsubsection{Collaboration Cost}
The collaborative authentication recruiting  more cooperative peers results in  higher cost. This is because longer  time and higher  overhead  will be required to request the collaboration and to achieve an agreement on the authentication decision. To achieve low collaboration cost, while maintaining high authentication accuracy, the optimal number  of cooperative peers will be determined.

\subsubsection{Authentication Robustness}
In the mobile network, the network topology could be rapidly fluctuating, where the neighbors of the user to be authenticated dynamically vary.   Hence, the  automatic update of the heterogeneous cooperative peers list and authentication features is extremely helpful for improving the authentication robustness with guaranteed decision accuracy.

\subsection{Problem Formulation}

According to the proposed authentication system
  relying on the cooperation of multiple edge devices, our problem
 can be formulated as follows.  Specifically, the objective
  function (OF) of collaborative authentication is denoted as
  $f(\bm{x}_{1},\bm{x}_{2},...,\bm{x}_{N})$ with respect to the local
  variables of $N$ cooperative peers.
 \begin{eqnarray}\nonumber
\min f(\bm{x}_{1},\bm{x}_{2},...,\bm{x}_{N}),~~~~\\
{\rm{s.t.}}~\bm{x}_{n}-\bm{z}=0,n=1,2,...,N,
\end{eqnarray}
where $\bm{x}_{n}=(x_{n1},x_{n2},x_{n3})^{\rm{\dagger}}$ are local variables and  $\bm{z}=(z_{1},z_{2},z_{3})^{\rm{\dagger}}\in\Re^{3}$ is the global variable.
Since the cooperative peers are at different locations and their observations are independent, the $f(\bm{x}_{1},\bm{x}_{2},...,\bm{x}_{N})$ is separable. Then,  we have
\begin{eqnarray}
f(\bm{x}_{1},\bm{x}_{2},...,\bm{x}_{N})=\sum_{n=1}^{N}f_{n}(\bm{x}_{n}),
\end{eqnarray}
where $f_{n}(\bm{x}_{n})$ is the OF of the $n$-th cooperative
peer. Upon denoting the real position of the $n$-th
  cooperative peer as
  $\bm{b}_{n}=(b_{n1},b_{n2},b_{n3})^{\rm{\dagger}}$, its local OF is
  given by
 \begin{eqnarray}
f_{n}(\bm{x}_{n})=
\begin{cases}
\parallel  \bm{x}_{n}- \bm{b}_{n}\parallel_{2}-H_{n1}~ &{\rm{RSSI}} \\
\parallel  \bm{x}_{n}- \bm{b}_{n}\parallel_{2}-H_{n2}~   &{\rm{TRA}}
\end{cases}.
\end{eqnarray}
It can be observed that the local OF is designed as a closed, proper, and convex function based on the observations of the user's features. $H_{n1}$ and $H_{n2}$ represent the $n$-th cooperative peer's observations of distance between the user to be authenticated and itself relying on the RSSI and TRA, respectively. Hence, $f_{n}(\bm{x}_{n})$ represents the difference between the local variable and real observation.
  Before detailing the proposed scheme further, we have provided a
  diagram in Fig. 2 for visualizing the detailed flow of our solution in the
  sequel.

\begin{figure}[htbp]
\centering
\includegraphics[width=8.5cm,height=12cm]{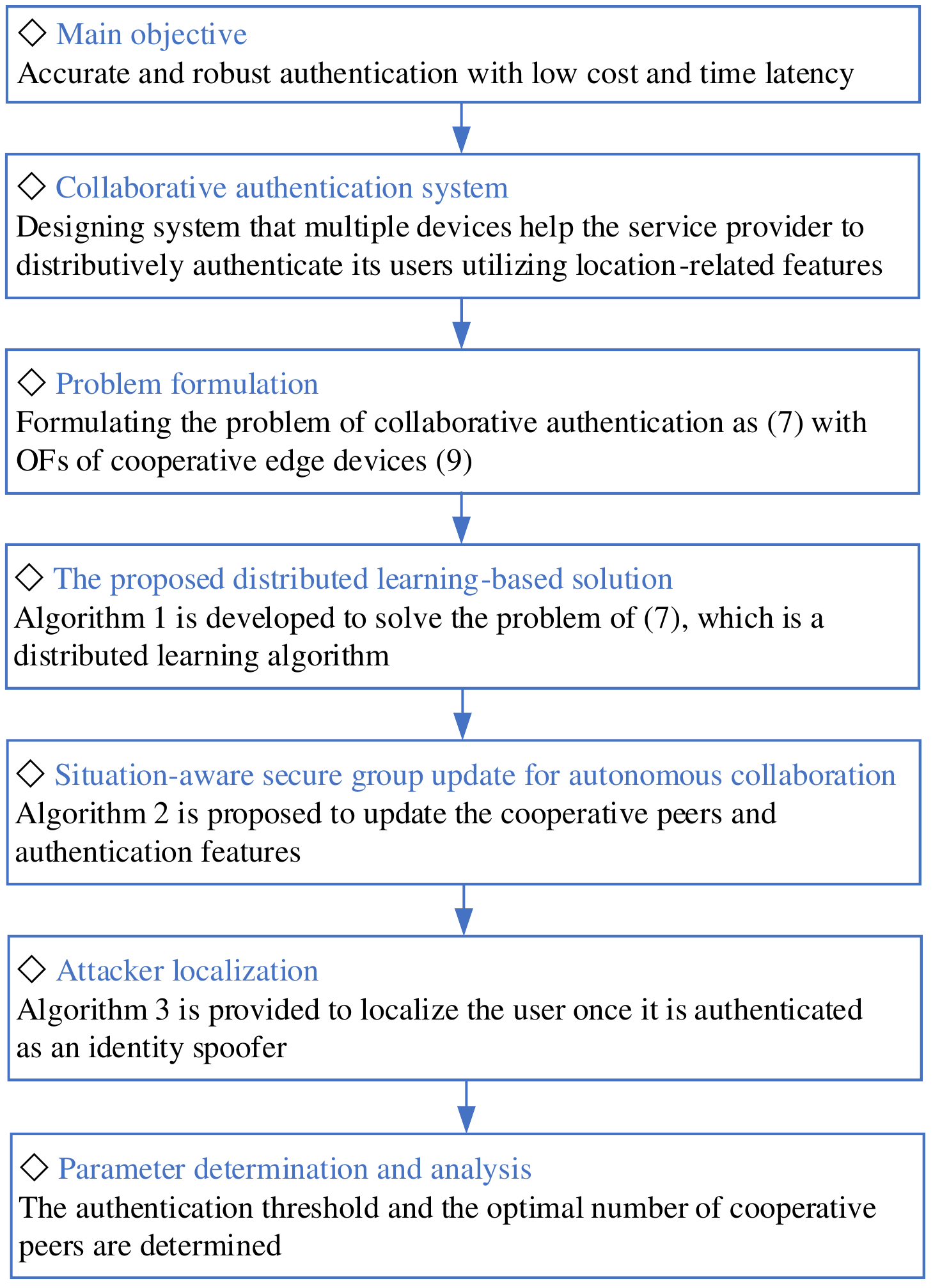}
\caption{Visualizing the proposed solution.}
\end{figure}

\section{EDGE INTELLIGENCE-BASED   AUTONOMOUS   COLLABORATIVE  AUTHENTICATION}
In achieving the accurate security provision, we propose an edge
intelligence-based collaborative authentication scheme. Firstly, a
distributed learning-based framework is developed, which relies on
training harvesting by multiple cooperative peers. An information
sharing and fusion strategy is designed in this framework to
amalgamate the information collected and to train the models at the
network edge, so that the network congestion and response
time can be reduced. Then, a situation-aware secure
  group update algorithm is developed for autonomously updating both the
  set of cooperative peers and the resultant features, so that the
  authentication robustness and reliability can be improved. Finally,
  an algorithm is proposed for localizing the attacker once it is identified.

\subsection{Distributed Learning-based Collaborative Authentication}

In order to solve the optimization problem of (7) associated with $N$ objective terms, the consensus alternating direction method of multipliers (ADMM) \cite{28} is applied. The  consensus ADMM algorithm for solving the problem of (7) can be derived  from the following augmented Lagrangian function
\begin{eqnarray}\nonumber
L_{n,\rho}(\bm{x}_{n},\bm{z},\bm{y}_{n})~~~~~~~~~~~~~~~~~~~~~~~~~~~~~~~~~~~\\
=f_{n}(\bm{x}_{n})+\bm{y}_{n}^{{\rm{\dagger}}}(\bm{x}_{n}-\bm{z})+(\rho/2)\|\bm{x}_{n}-\bm{z}\|_{2}^{2},
\end{eqnarray}
where $\rho>0$ is termed as the penalty parameter and $\bm{y}_{n}=(y_{n1},y_{n2},y_{n3})^{\rm{\dagger}}$.
Then, the consensus ADMM algorithm conceived for solving the  problem of (7) is given by
\begin{eqnarray}
&&\bm{x}_{n}^{k+1}:=\arg\min_{\bm{x}_{n}} L_{n,\rho}(\bm{x}_{n},\bm{z}^{k},\bm{y}_{n}^{k}),\\
&&\bm{z}^{k+1}:=\frac{1}{N} \sum_{n=1}^{N} (\bm{x}_{n}^{k+1}+(1/\rho)\bm{y}_{n}^{k}),\\
&&\bm{y}_{n}^{k+1}:=\bm{y}_{n}^{k}+\rho(\bm{x}_{n}^{k+1}-\bm{z}^{k+1}).
\end{eqnarray}

Specifically, we can observe from (9) and (10) that
  the expression $L_{n,\rho}(\bm{x}_{n},\bm{z},\bm{y}_{n})$ is
  a closed, proper and convex function. Hence, $\arg\min_{\bm{x}_{n}}
  L_{n,\rho}(\bm{x}_{n},\bm{z}^{k},\bm{y}_{n}^{k})$ can be readily
  solved \cite{80}.  Moreover, observed from (12) that the update of
$\bm{z}$ depends on the average values of $\bm{x}_{n}$ and
$\bm{y}_{n}, n=1,2,...,N$. Upon denoting them as $\overline{\bm{x}}$
and $\overline{\bm{y}}$, respectively, the learning process of (11)-(13) can be simplified as
\begin{eqnarray}
&&\bm{x}_{n}^{k+1}:=\arg\min_{\bm{x}_{n}} L_{n,\rho}(\bm{x}_{n},\overline{\bm{x}}^{k},\bm{y}_{n}^{k}),\\
&&\bm{y}_{n}^{k+1}:=\bm{y}_{n}^{k}+\rho(\bm{x}_{n}^{k+1}-\overline{\bm{x}}^{k+1}).
\end{eqnarray}

To achieve efficient collaborative authentication relying on multiple cooperative peers, a consensus is required, which contains the information sharing and fusion as well as the variable update in the proposed scheme:
1) To reduce the communication overhead, the collected authentication information and training models of cooperative peers will not be uploaded to the service provider;
2) As we can observe from (11)-(13), the $\bm{x}$-update and $\bm{y}$-update depend on the average values $\overline{\bm{x}}$ and $\overline{\bm{y}}$. Hence, these local parameters are shared among all the cooperative peers. The variables are separately updated to drive the variables towards consensus, and quadratic regularization helps pull the variables toward their average values, while still attempting to minimize each local function $f_{n}(\bm{x}_{n})$.

\begin{figure}[htbp]
\centering
\includegraphics[width=8cm,height=6cm]{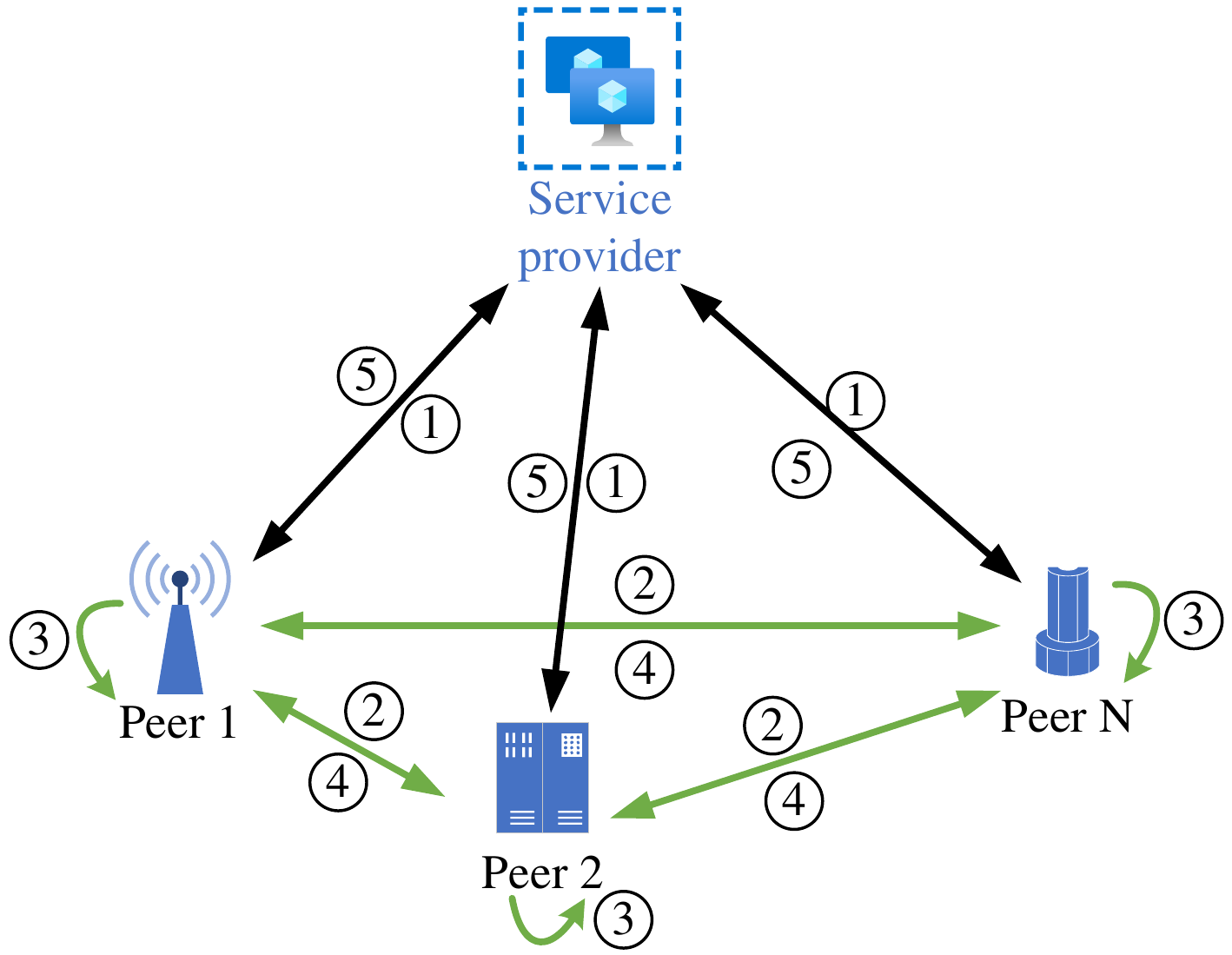}
\caption{Distributed learning-based collaborative  authentication framework.}
\end{figure}
As shown in Fig. 3, the distributed learning-based collaborative
 authentication framework contains the following steps:
 \\ \textit{Step $\textcircled{\small{1}}$.} The
   service provider sends a collaboration request to multiple edge
   nodes, which are located in the vicinity of the user to be
   authenticated. The available devices send
   their agreements to the service provider, which will act as cooperative peers;\\ \textit{Step
     \textcircled{\small{2}}.}  The associated cooperative peers estimate the
   features of the user to be authenticated, and share their estimates
   with other cooperative peers;\\ \textit{Step \textcircled{\small{3}}.} All the
 cooperative peers locally update their variables via
 (14)-(15);\\ \textit{Step \textcircled{\small{4}}.} All the
 cooperative peers share the learning parameters $\bm{x}_{n}^{k}$ with
 each other;\\ \textit{Step \textcircled{\small{5}}.}  If a convergence
 is achieved, i.e. $\bm{x}_{0}$, the authentication decision can be made at the
 cooperative peers via (4), and then they send their decisions to the
 service provider.

The proposed distributed learning-based solution  is formulated  in Algorithm 1.
\begin{algorithm}
    \caption{Distributed learning-based collaborative   authentication}
    \label{alg:Framwork}
    \begin{algorithmic}[1]
    \REQUIRE   $N$ cooperative peers, authentication threshold $\upsilon$, error bound of learning $\varrho$;
    \ENSURE  authentication results, i.e. legitimate user, identity spoofer, location spoofer;
       \STATE \textbf{for} each cooperative peer \textit{n} in parallel
       \STATE ~~~~estimates the authentication feature of user, i.e. RSSI \\
       ~~~~or TRA;
       \STATE ~~~~\textbf{if} the user cannot be observed
       \STATE ~~~~~~~~it is identified as a location spoofer;
       \STATE ~~~~\textbf{end if}
        \STATE ~~~~shares feature estimations with others;
        \STATE ~~~~\textbf{while} $\parallel\bm{x}_{n}^{k}-\bm{x}_{n}^{k-1}\parallel>\varrho$, \textbf{do}
         \STATE ~~~~~~~share parameters $\bm{x}_{n}^{k}$ with others;
       \STATE ~~~~~~~update local variables $\bm{x}_{n}$ and $\bm{y}_{n}$ via (14) and\\
        ~~~~~~~(15), respectively;
       \STATE  ~~~~\textbf{end}
         \STATE   \textbf{end for}
         \STATE  \textbf{if} a convergence  $\bm{x}_{0}$ is achieved  and $\parallel\bm{x}_{0}-\bm{a}\parallel_{2}\leq\nu$
          \STATE  ~~~~the user is authenticated as a legitimate device;
           \STATE  \textbf{otherwise}
            \STATE  ~~~~the user is identified as an identity spoofer;
             \STATE  \textbf{end if}
             \STATE  cooperative peers send the authentication decision to the
             service provider.
    \end{algorithmic}
\end{algorithm}

We can interpret the distributed learning-based solution of Algorithm 1 as a method for solving problems in which the objective and constraints are distributed across $N$ cooperative peers. Each cooperative peer only has to handle its own objective and constraint term, plus a quadratic term which is updated each iteration.
The quadratic terms (or more accurately, the linear parts of the quadratic terms) are updated in such a way that the variables converge
to a common value, i.e. $\bm{x}_{1}=\bm{x}_{2}=\cdots \bm{x}_{N}= \bm{z}$, which is the solution of (7).
Given the OF of each cooperative peer
  $f_{n}(\bm{x}_{n})$ in (9), we can see that they are closed,
  proper, and convex.  According to~\cite{28,32}, the proposed distributed learning-based algorithm
can   converge  to the real position of the user to be authenticated
  relying on its feature observations. The convergence process can be
expressed as
\begin{eqnarray}
\bm{x}_{n}^{k} \rightarrow  \bm{x}_{0}, n=1,2,...,N, {\rm{as}}~k \rightarrow \infty.
\end{eqnarray}
\textbf{Remark 1:} It can be observed from Algorithm 1 that the collaborative authentication depends on the consensus of the cooperative peers relying on their observations.  The user to be authenticated is only confirmed to be a legitimate one, when the convergence of the cooperative peers points to the real position of the legitimate user. Otherwise, it is deemed to be a spoofer. The proposed scheme moves the security provision to network  edge without involving  a centralized party. Moreover, the inherent features of the user to be authenticated are observed and exploited  by the cooperative peers to enhance the security. Hence, it is extremely difficult for the attacker to crack the authentication method by imitating all the features as well as by misleading  all the cooperative peers which  are located near to the user.

\subsection{Situation-Aware Autonomous Collaboration}
To meet the stringent 6G networking demands, heterogeneous devices with  time-varying position, network connection, and power/battery status will have to be supported in complex dynamic environments. To be more specific,
when the user is moving, its cooperative peers may lose connection with it. Furthermore, due to the resource limitation of cooperative peers, e.g. battery and storage, some of them may be unavailable for helping the service provider after a while. To achieve flexible and robust collaborative security provision, a situation-aware secure group update algorithm is developed in Algorithm 2 to achieve  autonomous collaborative authentication by adaptively updating the cooperative peers list and the associated authentication features.

Upon denoting the initial set of cooperative peers as $\Lambda[0]$, Algorithm 2 updates $\Lambda[t]$ by adaptively identifying the dynamic environment and network topology to achieve reliable and robust collaborative authentication at time instance $t$. Specifically, when one of the cooperative peers leaves set $\Lambda[t]$, another edge node  will be asked to join in the collaborative authentication. The update of secure group is shown as
\begin{eqnarray}
\Lambda[t+1]=\Lambda[t]-\{\widehat{\gamma}[t]\}+\{\gamma[t]\},
\end{eqnarray}
where $\widehat{\gamma}[t]$ is the cooperative peer left the secure group at time   $t$. The selection of new cooperative peer is formulated as
\begin{eqnarray}
\gamma[t]=\arg\min_{j\in\Upsilon} \|\bm{x}_{0}[t]-\widehat{\bm{b}}_{j}\|_{2},
\end{eqnarray}
which represents the available edge node located nearest to the user to be authenticated. $\Upsilon$ is the set of edge nodes, and    $\widehat{\bm{b}}_{j}=(\widehat{b}_{j1},\widehat{b}_{j2},\widehat{b}_{j3})^{\rm{\dagger}}$ denotes the real position of the $j$-th edge node, $j\in \Upsilon$.

\begin{algorithm}
    \caption{Situation-aware secure group update}
     \label{alg:Framwork}
    \begin{algorithmic}[1]
    \REQUIRE   initial set of cooperative peers $\Lambda[0]$, set of edge nodes $\Upsilon$ and their real positions $\widehat{\bm{b}}_{j}$;
    \ENSURE  updated set of cooperative peers $\Lambda[t+1]$ and authentication result;
     \STATE \textbf{for} authentication round $t=1,2,3,...$
      \STATE  ~~~\textbf{if} a cooperative peer leaves the $\Lambda[t]$
      \STATE  ~~~~~~~~the service provider sends collaboration request \\
     ~~~~~~~~to more devices located near the user to be \\
      ~~~~~~~~authenticated;
       \STATE  ~~~~~~~~available devices send  feedback to the service\\
       ~~~~~~~~provider;
       \STATE  ~~~~~~~~the set of cooperative peers is updated via (17);
     \STATE ~~~\textbf{end if}
         \STATE  ~~~all cooperative peers perform Algorithm 1 and send \\
         ~~~their availability for the next round of authentication\\
         ~~~to the service provider;
         \STATE \textbf{end for}
    \end{algorithmic}
\end{algorithm}

It can be observed from Algorithm 2 that efficient authentication can be achieved through autonomous update of cooperative edge nodes and features. To be more specific, the service provider can supply a list of volunteering  cooperative peers ahead of time, so that the latency of collaborative authentication can be reduced. Note that the proposed scheme will be suitable for those communication systems with sufficient edge nodes, e.g. intelligent building, Internet of Vehicles (IoV), VANETs, AANETs, and UAV networks, just to name a few.

\subsection{Attacker Localization}
Once an attacker, who performs identity spoofing attacks,  is detected by the proposed collaborative authentication scheme via  (4),
it can also be localized by Algorithm 3.  Upon denoting the real position of this attacker by $\bm{c}=(c_{1},c_{2},c_{3})^{\rm{\dagger}}$, the $N$ cooperative peers can observe the location-related  features of this attacker, i.e. its  RSSI and TRA. Then,
the convergence result of Algorithm 1 relying on the attacker's feature estimation can be formulated as
\begin{eqnarray}
\bm{x}_{n}^{k} \rightarrow \bm{x}_{0}=\bm{c}+\bm{\varepsilon}, n=1,2,...,N, {\rm{as}}~k \rightarrow \infty,
\end{eqnarray}
where $\bm{\varepsilon}$ is the learning error due to the imperfect estimations of features used in the proposed scheme.

\begin{algorithm}
    \caption{Attacker localization }
        \label{alg:Framwork}
    \begin{algorithmic}[1]
    \REQUIRE  result of Algorithm 1;
    \ENSURE  location of the identity spoofer;
     \STATE \textbf{if} the user is authenticated as an identity spoofer
      \STATE ~~~\textbf{if} a consensus $\bm{x}_{0}$ is achieved among $N$ peers
      \STATE ~~~~~~~the attacker can be  localized as $\bm{x}_{0}$ of (19);
      \STATE ~~~\textbf{end if}
      \STATE \textbf{end if}
    \end{algorithmic}
\end{algorithm}

Note that the attacker cannot perfectly imitate the legitimate user and cheat  all the cooperative peers when it locates at a different position from the legitimate user, especially when the proposed  scheme utilizes more cooperative peers. This is not unexpected,  because the cooperative peers locate near to the user to be authenticated and can observe different features as well as other information, e.g. certificate. It becomes more and more difficult for the attacker to spoof the legitimate user, when the number of cooperative peers is increased, demonstrating the benefit of the proposed scheme.\\
\textbf{Remark 2:} The attacker can only be localized, if it is an identity spoofer and at least one of the cooperative peers can identify the real ID of the attacker. If the attacker aims for performing  location spoofing attacks, it cannot be localized  by the cooperative peers, but it will be identified as a location spoofer directly. Moreover, once one of the cooperative peers is deceived  by the attacker, the proposed collaborative authentication solution (i.e. Algorithm 1) will be divergent. In this case, the user to be authenticated will also be deemed to be an attacker.

\subsection{Parameter Determination and Theoretical Analysis}
In order to achieve improved  authentication performance, including higher authentication accuracy and reduced  collaboration cost,  the authentication threshold  $\nu$ and the number of cooperative peers $N$ are determined in this subsection.

\subsubsection{Determination of the threshold $\nu$}
As it is shown in (4), the authentication accuracy depends on the
threshold $\nu$. A lower threshold $\nu$ will lead to higher FA rate, while a higher $\nu$ will result in higher MD
rate. Hence, there is a trade-off between the
 MD rate and FA rate with respect to
  authentication threshold $\nu$. In order to achieve a better
security performance, we formulate the threshold determination problem
as
\begin{eqnarray}
\min_{\nu} F_{\rm{FA}},~~~\\\nonumber
{\rm{s.t.}} ~F_{\rm{MD}}<\epsilon,
\end{eqnarray}
where $\epsilon$ is the constraint of MD rate.

In this subsection, we consider an example, where
the potential distances between the attacker and the legitimate user (denoted as $d_{a}$) obey the Log-normal distribution, which satisfies
\begin{eqnarray}
\ln d_{a} \sim \aleph (\mu_{a},\sigma_{a}^{2}),
\end{eqnarray}
where $\mu_{a}$ and $\sigma_{a}^{2}$ represent the mean and variance of the Gaussian distribution $\ln d_{a}$, respectively. $\aleph$ denotes the Gaussian distribution symbol. Then, the probability density function (PDF) of the above log-normal distribution is given by
\begin{eqnarray}
g_{\Im}(\varsigma)=\frac{1}{\sigma_{a}\varsigma\sqrt{2\pi}}e^{-\frac{(\ln \varsigma-\mu_{a})^{2}}{2\sigma_{a}^{2}}},
\end{eqnarray}
where its mean and variance are
  $e^{\mu_{a}+\sigma_{a}^{2}/2}$ and
  $(e^{\sigma_{a}^{2}}-1)e^{2\mu_{a}+\sigma_{a}^{2}}$,
  respectively.\\
  \textbf{Remark 3:} It is
  reasonable to assume that the potential distances between the
  attacker and user obey Log-normal distribution, since a smart
  attacker focuses on observing and imitating the legitimate users, but
  it wants to avoid being spotted by the user at the same time. Hence,
  to glean increased utility, a spoofer will opt for a position
  having an appropriate distance from the user to be authenticated.
  Note that it is not critical to let the distances obey the
  Log-normal distribution - we only consider a reasonable example to
  provide an expression for the authentication threshold $\nu$ for
  the analysis in this section. In a specific communication application
  scenario, the distance distribution can be obtained according to the
  historical knowledge of the attacker detection by the service provider.

Given a specific application scenario, e.g. VANET and UAV network,  if there is no knowledge of $\mu_{a}$ and $\sigma_{a}^{2}$, the initial authentication threshold is set to
\begin{eqnarray}
\nu_{\rm{Ini}}=\iota_{0}+\iota,
\end{eqnarray}
where $\iota_{0}$ and $\iota$ denote the error value of the positioning technique used in the system and the detection error value of Algorithm 1  in [m], respectively.
In the other cases, the knowledge of $\mu_{a}$ and $\sigma_{a}^{2}$ could be inferred from  the historical information of attackers detected by the service provider, e.g. the location information of attacks obtained by Algorithm 3. Then, the following Theorem can be formulated.\\
\textbf{Theorem 1:} Given $\mu_{a}$ and $\sigma_{a}^{2}$, the authentication threshold $\nu$ of the proposed  scheme can be expressed   as
\begin{eqnarray}
\nu= \min\{\nu_{\rm{opt}},\nu_{\rm{Ini}}\},
\end{eqnarray}
where $\nu_{\rm{opt}}$ satisfies
\begin{eqnarray}
erf(\frac{\ln \nu_{\rm{opt}}-\mu_{a}}{\sqrt{2}\sigma_{a}})=2\epsilon-1.
\end{eqnarray}
$erf(\cdot)$ is the error function. \\
\textit{Proof:} Please see Appendix 1.\\
Then, the FA rate of the proposed scheme can be calculated via (6), which depends on the number of cooperative peers $N$ and their feature estimates.
\subsubsection{Determination of  number of cooperative peers $N$}
 The performance of  collaborative authentication depends on the features estimated by the cooperative peers. Due to their imperfect estimations, utilizing more cooperative peers in the proposed scheme  will reduce  the learning loss and increase the authentication accuracy, but will impose  high communication overhead and collaboration cost as well. Hence, there is a trade-off between the authentication performance and collaboration cost.
To achieve the best performance, an optimal number of peers can be derived by maximizing the authentication  accuracy  under a given maximum collaboration  cost threshold.
Let us denote the number of iterations for one round of authentication in Algorithm 1 by $K$,  the number of communication instances  among cooperative peers for one round  of collaborative authentication can be expressed as
\begin{eqnarray}
\varphi =(K+1)N(N-1).
\end{eqnarray}
Then, the following theorem can be formulated. \\
\textbf{Theorem 2:} Given the collaboration cost constraint $\tau$, the number of cooperative peers at time $t$ is determined as
\begin{eqnarray}
N [t] =\max \{N_{{\rm{min}}},\min \{N_{{\rm{opt}}}, N_{0}[t]\}\},
\end{eqnarray}
where $N_{{\rm{min}}} =3$ is the minimum number of cooperative peers.
$N_{0}[t]$ is the maximum number of devices that are available to help the service provider at time instant $t$.
The optimal  number of cooperative peers is given by
\begin{eqnarray}
N_{{\rm{opt}}} =  [\sqrt{\frac{\tau}{K_{{\rm{ave}}}+1}+\frac{1}{4}}+\frac{1}{2}],
\end{eqnarray}
where $[\cdot]$ is the least integer function. $K_{{\rm{ave}}}$ represents the average learning steps of one-round of collaborative authentication by Algorithm 1. \\
\textit{Proof:} Please see Appendix 2.\\
Note that the $K_{{\rm{ave}}}$ depends on the error bound of learning pre-set in Algorithm 1, i.e. $\varrho$.
Given the above choice of parameters, improved security and reduced cost can be achieved based on the proposed scheme.\\
\textbf{Remark 4:} Corresponding to the given criteria in Section II-C, the determination of the threshold $\upsilon$ relies on the authentication accuracy, and the determination of $N$ depends on the collaboration cost. To be more specific, there exists a trade-off between FA rate and MD rate with respect to threshold $\upsilon$. We determine the  threshold $\upsilon$  to achieve best authentication accuracy by solving the optimization problem of (20). Moreover, utilizing more cooperation peers will achieve better collaborative authentication accuracy while leading to higher cost. Hence, the determination of $N$ is to achieve best security performance under the collaboration cost constraint.

\section{PERFORMANCE ANALYSIS AND EVALUATION}
In evaluating the proposed scheme, two simulation examples are considered  in this section, namely an  indoor and an outdoor communication scenario. The results demonstrate the viability of the proposed scheme in the practical communication systems, such as the intelligent building, IoV,  UAV networks, and so on.

\subsection{Indoor Communication Scenario}
In our first example,  a $10\times 10$ m indoor office is simulated, where the user to be authenticated  is  located at position [6,6] m. The service provider is located outside this office.
5 devices are located randomly in this indoor office acting as cooperative peers by harnessing the  RSSI  and  TRA of the user to be authenticated for collaborative authentication.

\begin{figure}[htbp]
\centering
\includegraphics[width=8.5cm,height=7cm]{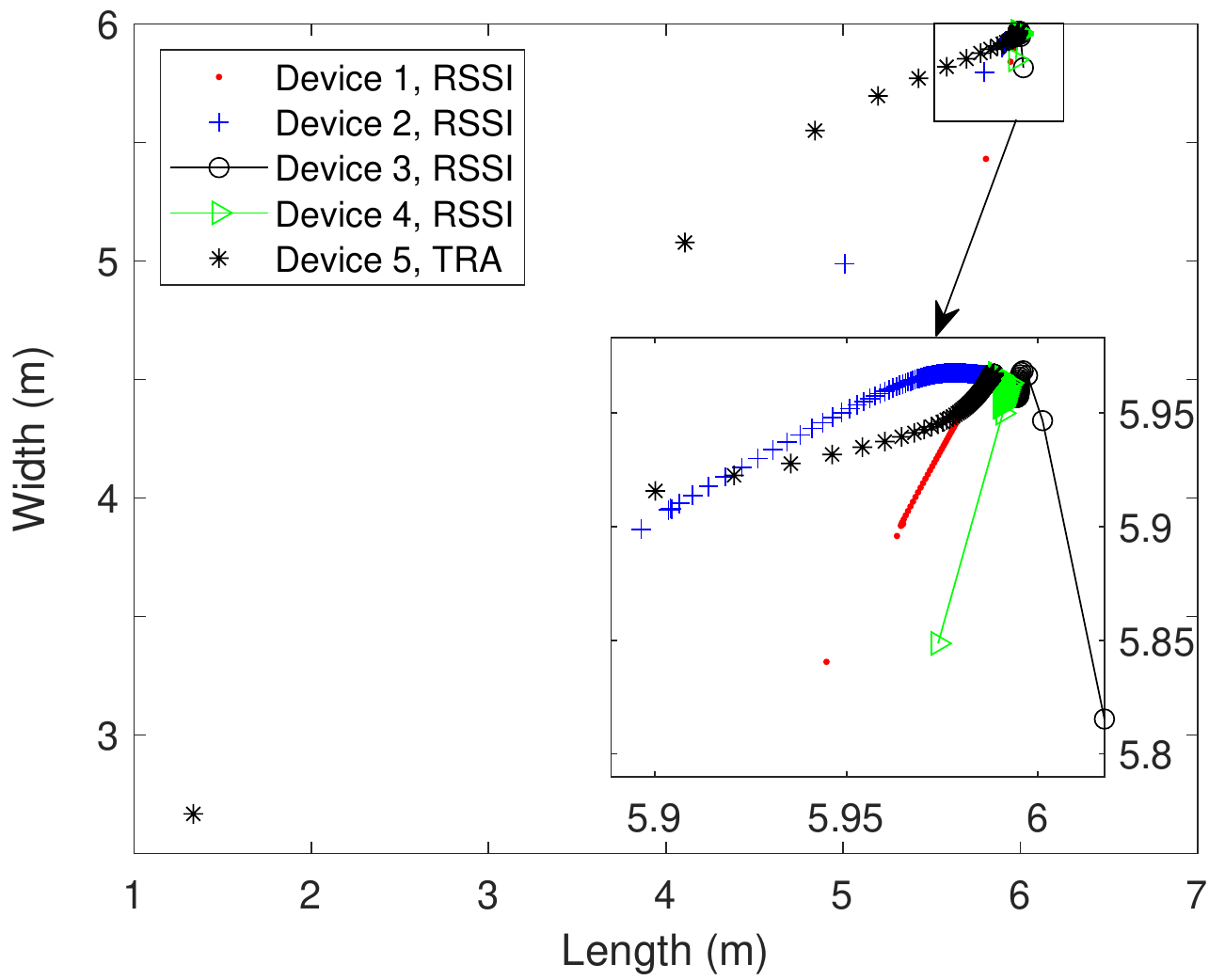}
\caption{Learning processes at the cooperative peers and the convergence of the proposed Algorithm 1  in an indoor communications scenario relying on 5 cooperative peers.}
\end{figure}

As shown in Fig. 4, Devices 1-4 use the RSSI and Device 5 utilizes the TRA for collaborative authentication, respectively. x-axis and y-axis represent the length and width of this indoor office in [m].
In this simulation, the service provider only has to know that Devices 1-5 are located in this office, so that the privacy of devices, including  their location and  mobility trajectory, can be concealed from the service provider while Devices 1-5 help the service provider to authenticate its user.
One can observe from Fig. 4 that all the learning parameters of Devices 1-5  converge to the real position of the legitimate user (i.e.  [6,6] m) based on Algorithm 1 relying on their local  RSSI and TRA information of the user. Hence, the user to be authenticated is identified as the legitimate user by the proposed scheme. This figure demonstrates the convergence results of Algorithm 1 and validate the proposed solution in collaboratively authenticating the users in a simulated indoor communication scenario.

\begin{figure}[htbp]
\centering
\includegraphics[width=8.5cm,height=7cm]{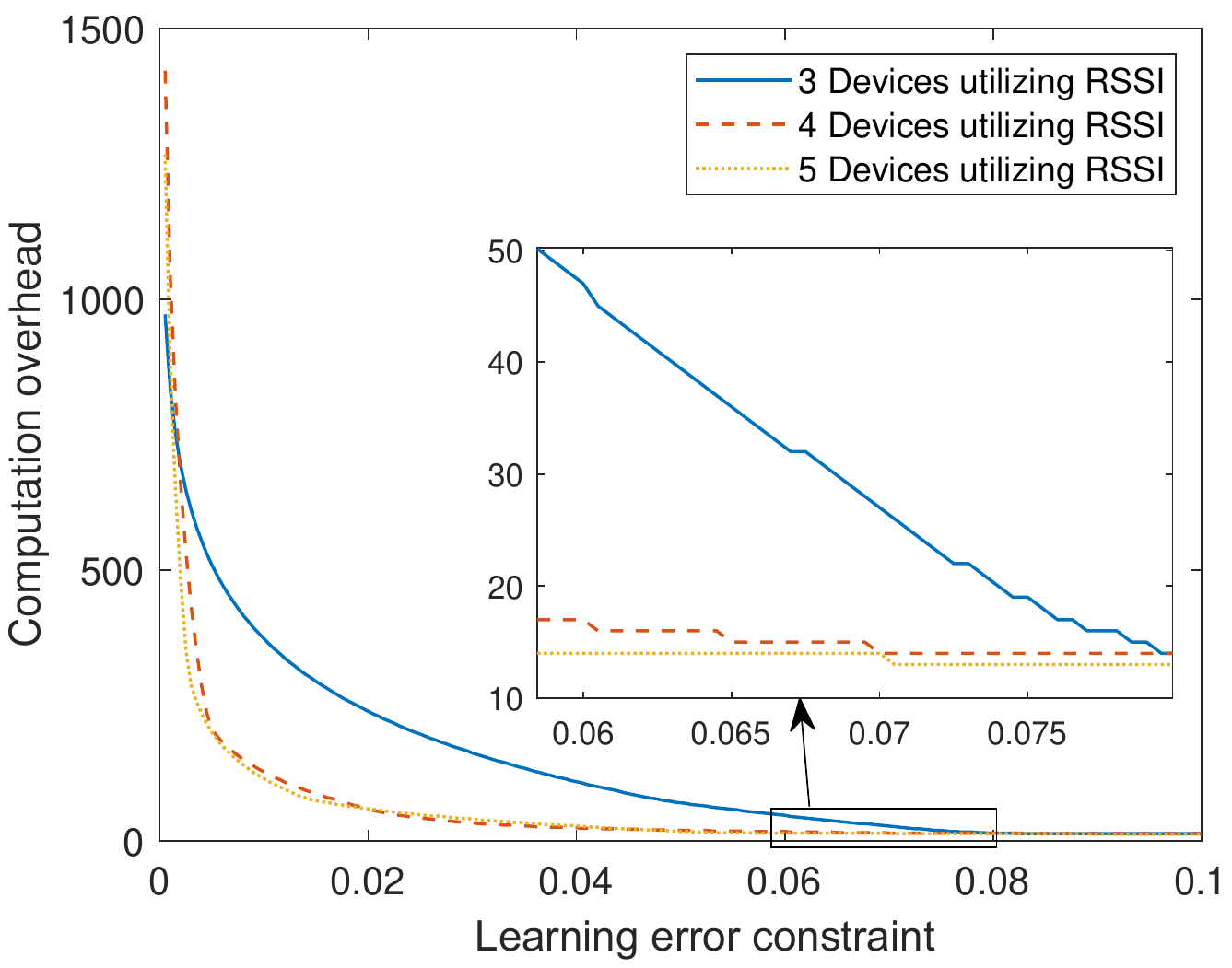}
\caption{Computation overhead of the proposed scheme with different numbers of cooperative peers utilizing RSSI.}
\end{figure}

Fig. 5 characterizes the computation overhead comparison results of the proposed scheme (i.e. Algorithm 1) relying on different numbers of cooperative peers utilizing the RSSI. We can observe from this figure that  the proposed scheme requires lower computation overhead to achieve a consensus at a given error constraint by utilizing more cooperative devices for collaborative authentication. The reason for this trend is that more cooperative peers used can achieve better learning  accuracy of Algorithm 1.
Moreover, the number of  computations required  for achieving convergence decreases upon increasing learning error constraint. It also exhibits the trade-off between the learning accuracy and computation overhead in the proposed distributed learning-based solution (i.e. Algorithm 1).

\begin{figure}[htbp]
\centering
\includegraphics[width=8.5cm,height=7cm]{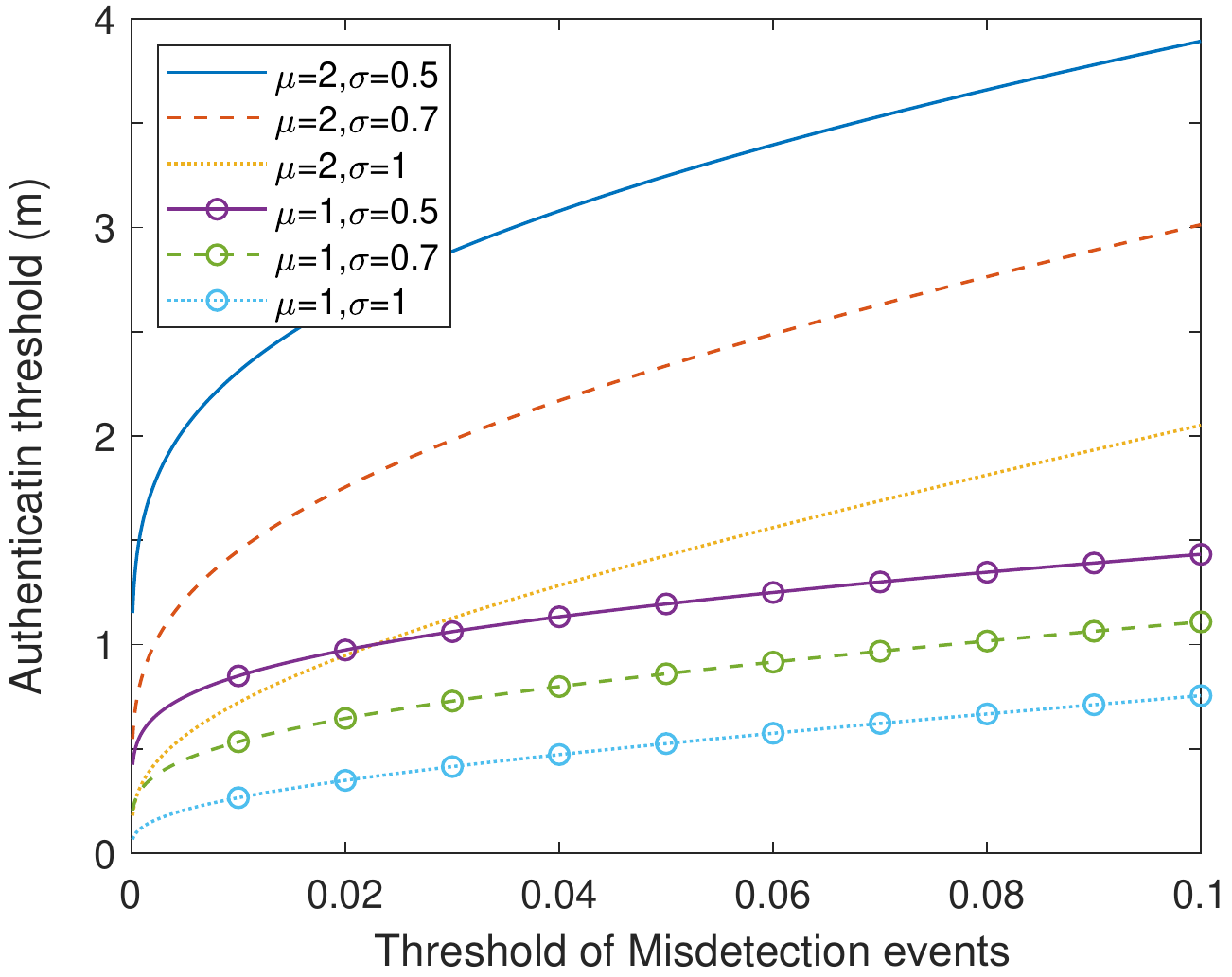}
\caption{Determined authentication threshold vs. threshold of MD rate of the proposed scheme in cases with different $\mu$ and $\sigma$ values of Log-normal distribution in (21).}
\end{figure}

The relationship between the
  threshold and authentication accuracy is characterized in Fig. 6. In
  this simulation, we assume that the potential distances between the
  attacker and the legitimate user obey the Log-normal distribution
  associated with different $\mu$ and $\sigma$ values.  Given the
  threshold of MD alert, the authentication threshold $\nu$
  of the proposed scheme can be determined based on the result of
  Theorem 1. We can observe from Fig. 6 that the increased threshold
  of MD rate will result in an increased authentication  threshold
  $\nu$. It is also shown in Fig. 6 that a smaller $\mu$ results in
  reduced authentication threshold $\nu$, because the
  attacker is located closer to the legitimate user in these cases.

\begin{figure}[htbp]
\centering
\includegraphics[width=8.5cm,height=7cm]{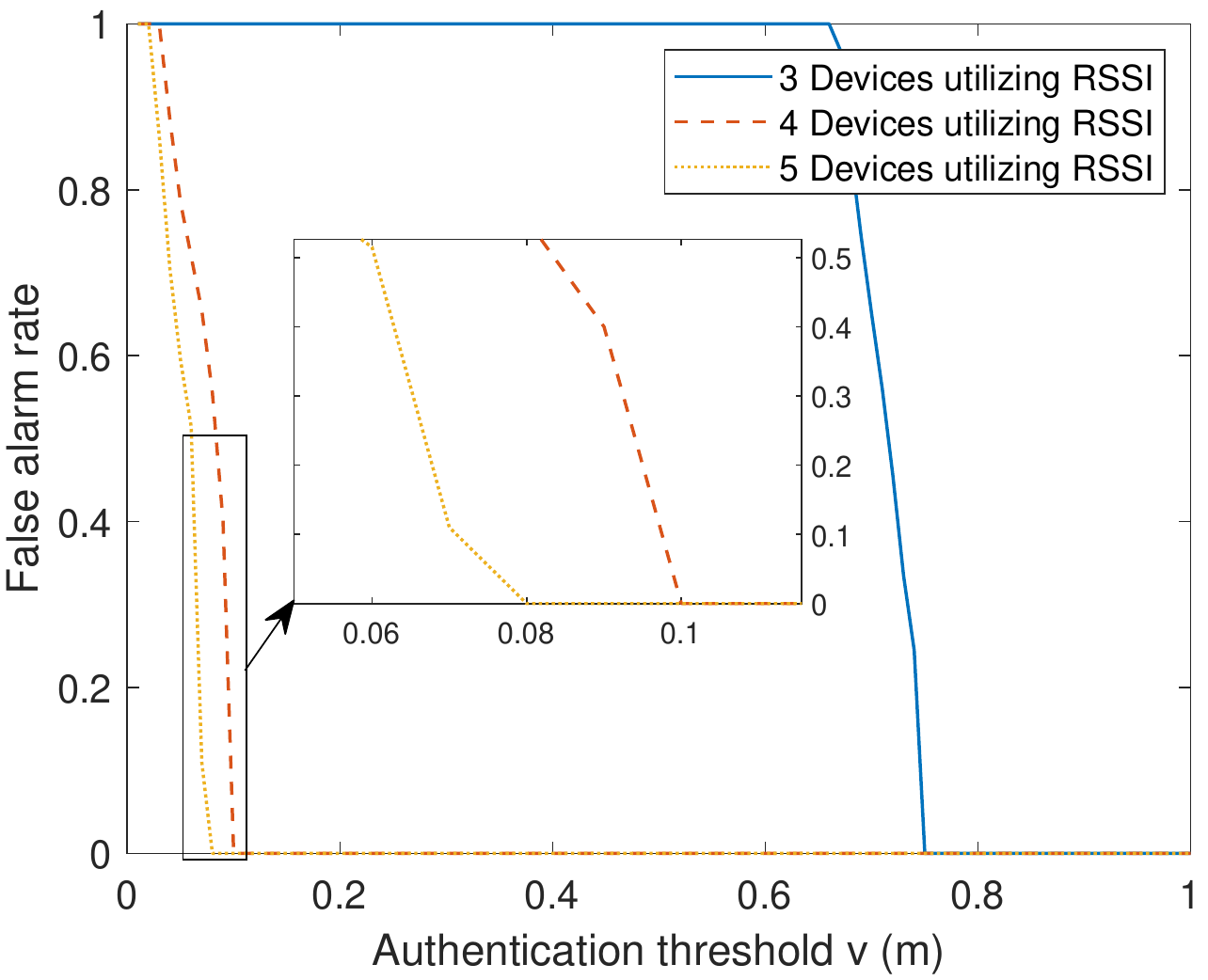}
\caption{Authentication threshold vs. FA rate of the proposed scheme with different numbers of cooperative peers.}
\end{figure}

Fig. 7 characterizes the authentication threshold  vs. the FA rate of the proposed scheme by using different numbers of cooperative peers. They estimate the RSSI of the user to be authenticated for verification. We can observe from Fig. 7 that the FA rate dramatically decreases  upon increasing the  authentication threshold $\nu$. This is because that the FA rate of the proposed scheme depends on the learning accuracy of the proposed distributed learning-based solution (i.e. Algorithm 1) relying on the imperfect RSSI estimation. Moreover, harnessing more cooperating peers reduces the FA rate of the proposed scheme.
Fig. 6 and Fig. 7 also demonstrate  that there is a trade-off between the FA rate and MD rate of the proposed scheme.
We can also observe from Fig. 5 and Fig. 7 that   the performance of the proposed scheme utilizing 4 cooperative devices is very close to that of utilizing 5  devices. Hence, we can choose 4 devices to be cooperators to reduce the collaboration cost in this case.

\subsection{Outdoor Communication Scenario}
In this subsection, an outdoor communication scenario
  is simulated, where the legitimate user's position is [0,0] m and its velocity is 10km/h, the identity attacker's position is [0,10] m.
   Fig.~8 characterizes the learning processes of cooperative peers and the
  convergence of the proposed scheme. It can be observed from this figure
  that the learning processes converge to [0.1,0.35] m, which is
  close to the user, i.e. [0,0] m. The detection error value in this
  case is 0.364 m, which is the distance between the convergence
  result of Algorithm 1 and the real position of the user. Given the authentication
threshold $\nu$ by Theorem 1, the user to be authenticated will be
identified as a legitimate user.

\begin{figure}[htbp]
\centering
\includegraphics[width=8.5cm,height=7cm]{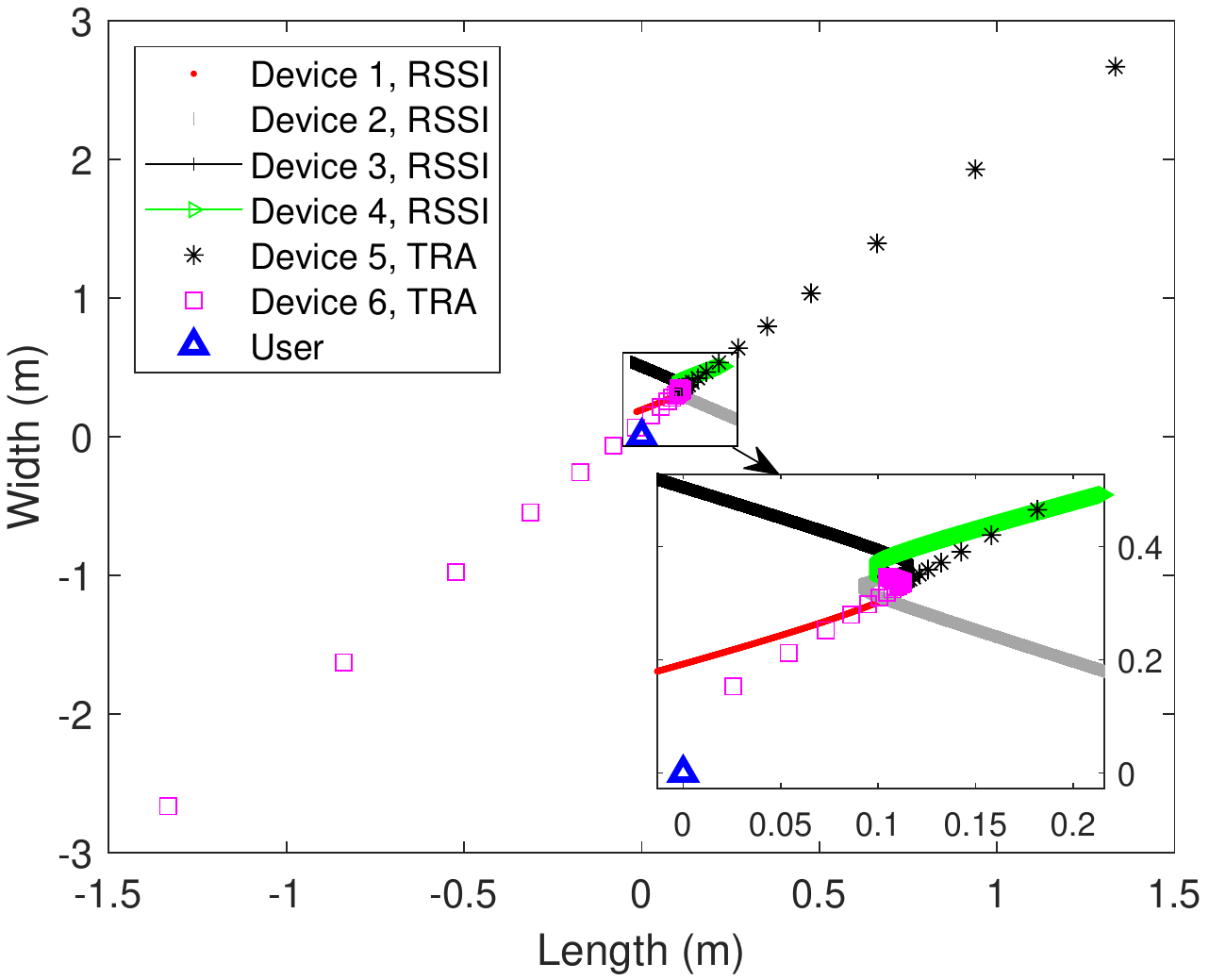}
\caption{Learning processes at cooperative peers  and the convergence of the proposed scheme in an outdoor communication scenario, where Devices 1-4 choose the RSSI and Devices 5-6 utilize TRA.}
\end{figure}

\begin{figure}[htbp]
\centering
\includegraphics[width=8.5cm,height=7cm]{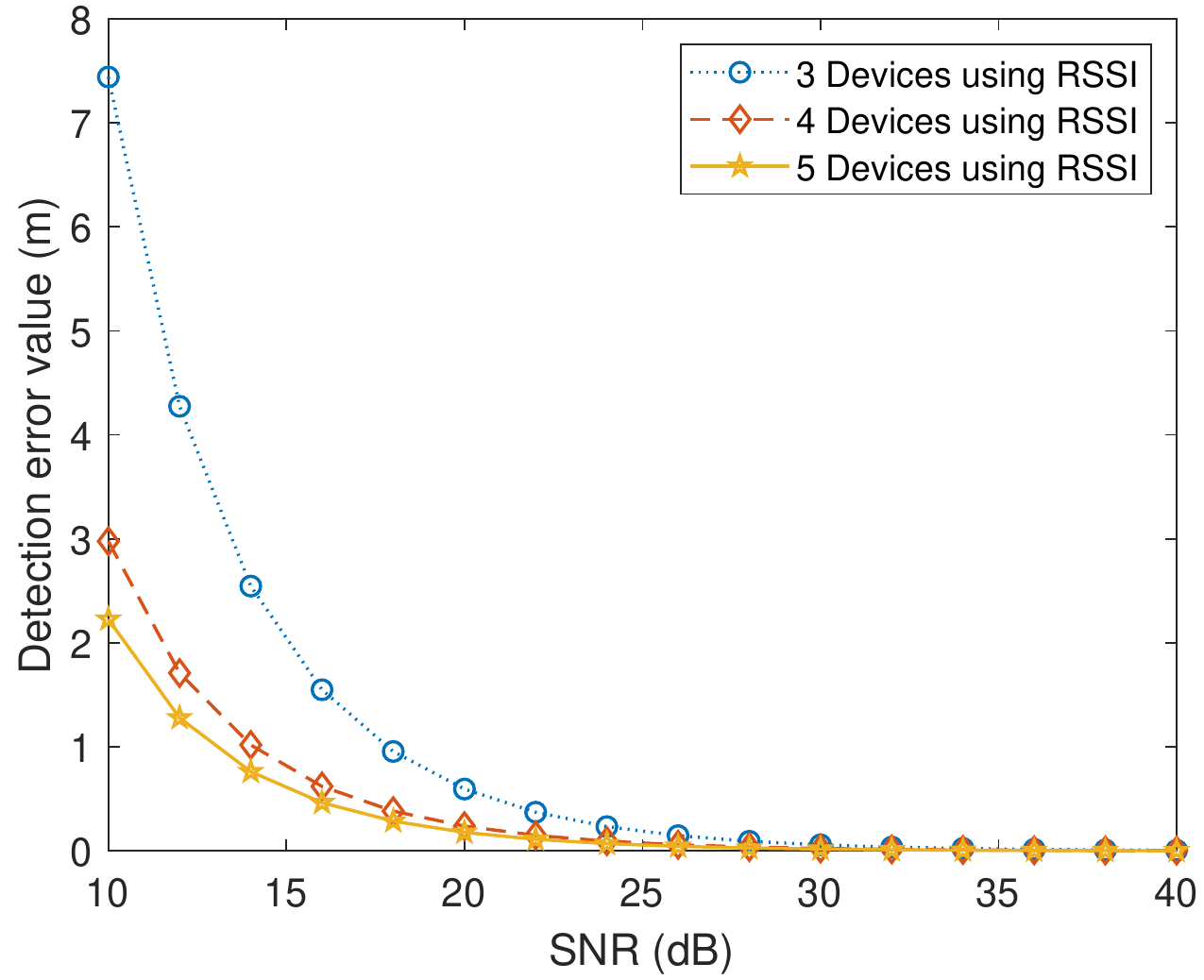}
\caption{SNR vs. detection error value of the proposed scheme in the cases that different numbers of cooperative peers are utilized, i.e. 5, 4, and 3.}
\end{figure}

Fig. 9 characterizes the signal-to-noise ratio (SNR) vs. the
authentication accuracy of the proposed scheme. The cooperative peers' positions  are   [-16,12] m, [-22,-20] m, [28,16] m, [24,-18] m, and [12,14] m.  We can observe from Fig. 9
that the detection error value  decreases
dramatically with the increase of SNR value, because the estimation
errors of RSSI are lower. The performance of the case that 5 devices using RSSI is the best. The reason is that more cooperative peers utilized in the proposed scheme can collect more RSSI information for collaborative authentication, and for compensating the estimation errors of RSSI to achieve better performance.

\begin{figure}[htbp]
\centering
\includegraphics[width=8.5cm,height=7cm]{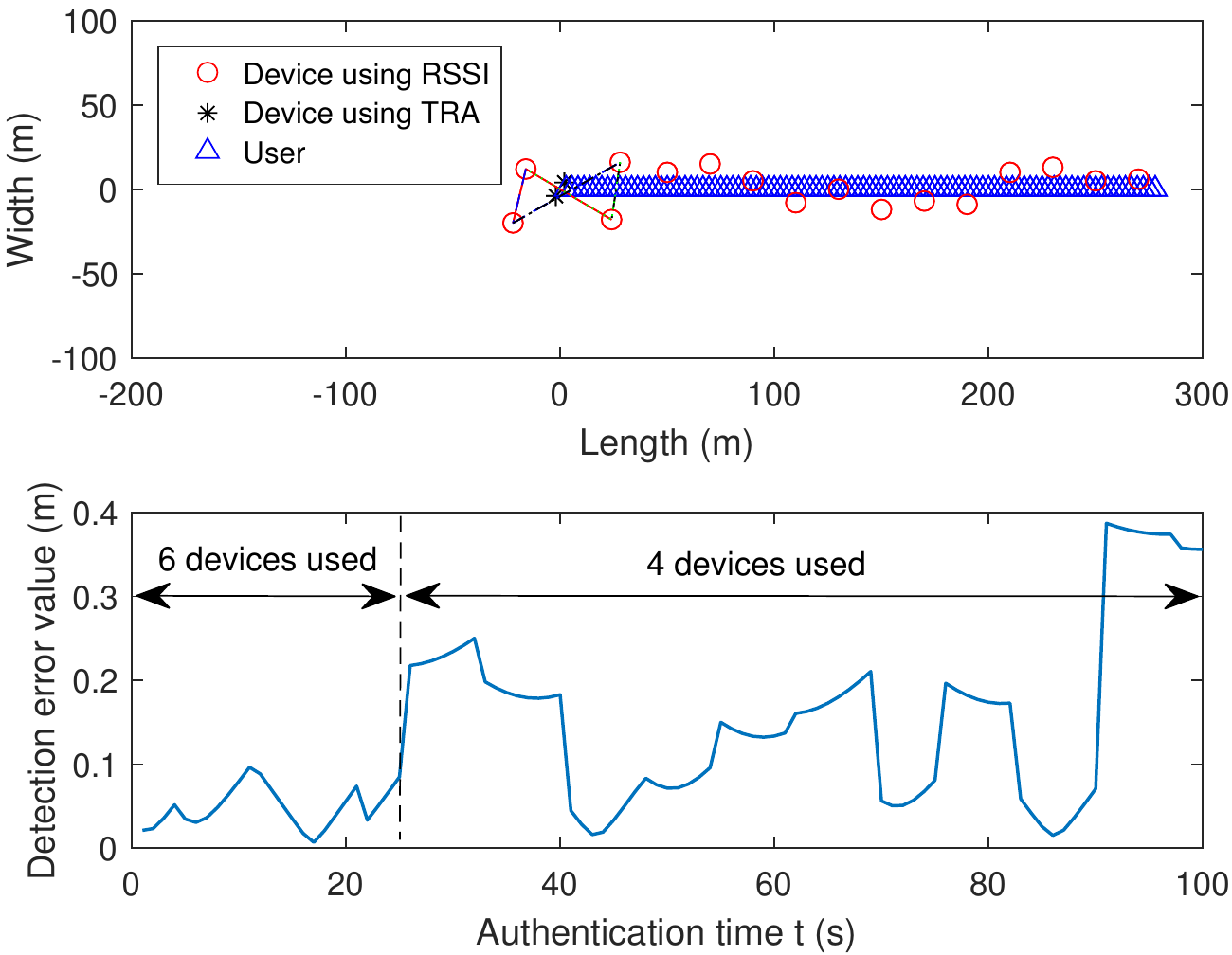}
\caption{Autonomous collaborative authentication process of the proposed scheme, where the  SNR  is 30dB.}
\end{figure}

Fig. 10 characterizes the simulated system topology relying on our
autonomous collaborative authentication process, where 6 cooperative
edge devices collected the observations of the user at
beginning. During the movement of the user, the proposed scheme
autonomously updates the cooperative devices.  In this simulation, the
other edge devices are located at [50,10] m, [70,15] m, [90,5] m,
[110,-8] m, [130,0] m, [150,-12] m [170,-7] m, [190,-9] m, [210,10] m,
[230,13] m, [250,5] m, and [270,6] m, which are shown as red circles
in Fig. 10.  In order to reduce the collaboration cost, the proposed
scheme utilizes 4 cooperative edge devices for collaborative
authentication based on situation awareness after the 25th
authentication episodes.   It can be observed from
  Fig.~10 that more devices autonomously join the collaborative
  authentication process utilizing the RSSI, and the proposed scheme
  continues operating during the user's movement. The results of
  Fig.~10 verify the authentication robustness of the proposed scheme
  in the noisy communication environment considered.

\begin{figure}[htbp]
\centering
\includegraphics[width=8.5cm,height=7cm]{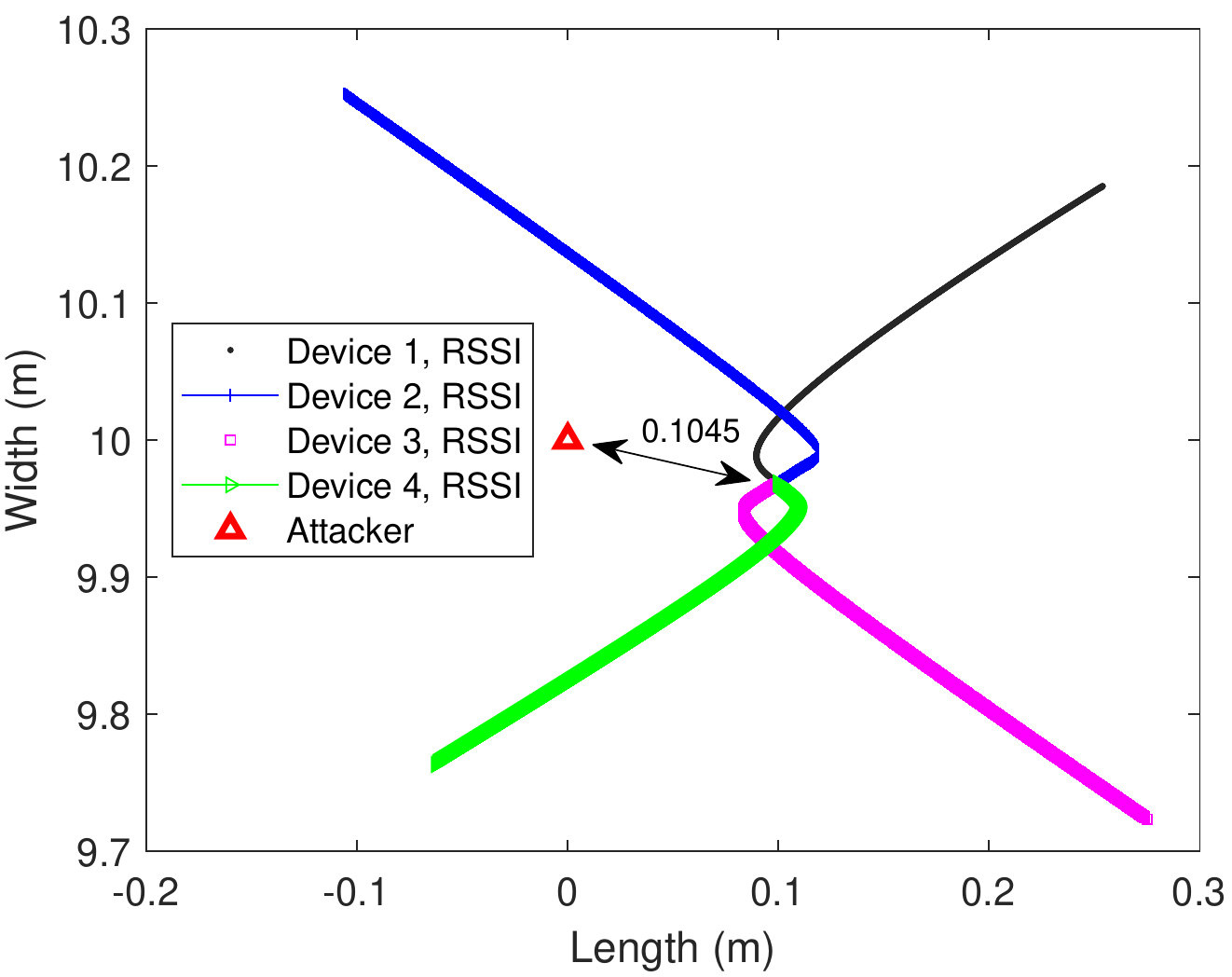}
\caption{Attacker localization based on the proposed scheme.}
\end{figure}

Fig.~11 characterizes the attacker localization
  process based on   Algorithm 3. In this
  simulation, 4 cooperative peers are utilized for collaborative
  authentication relying on the RSSI. It is observed from this figure that
  the learning processes converge to [0.1,9.96] m, which is close to
  the real position of the attacker, i.e. [0,10] m. The detection
  error value in this case is 0.1045 m. Hence, the location of the identity spoofer can be determined by the proposed scheme.

\begin{figure}[htbp]
\centering
\includegraphics[width=8.5cm,height=7cm]{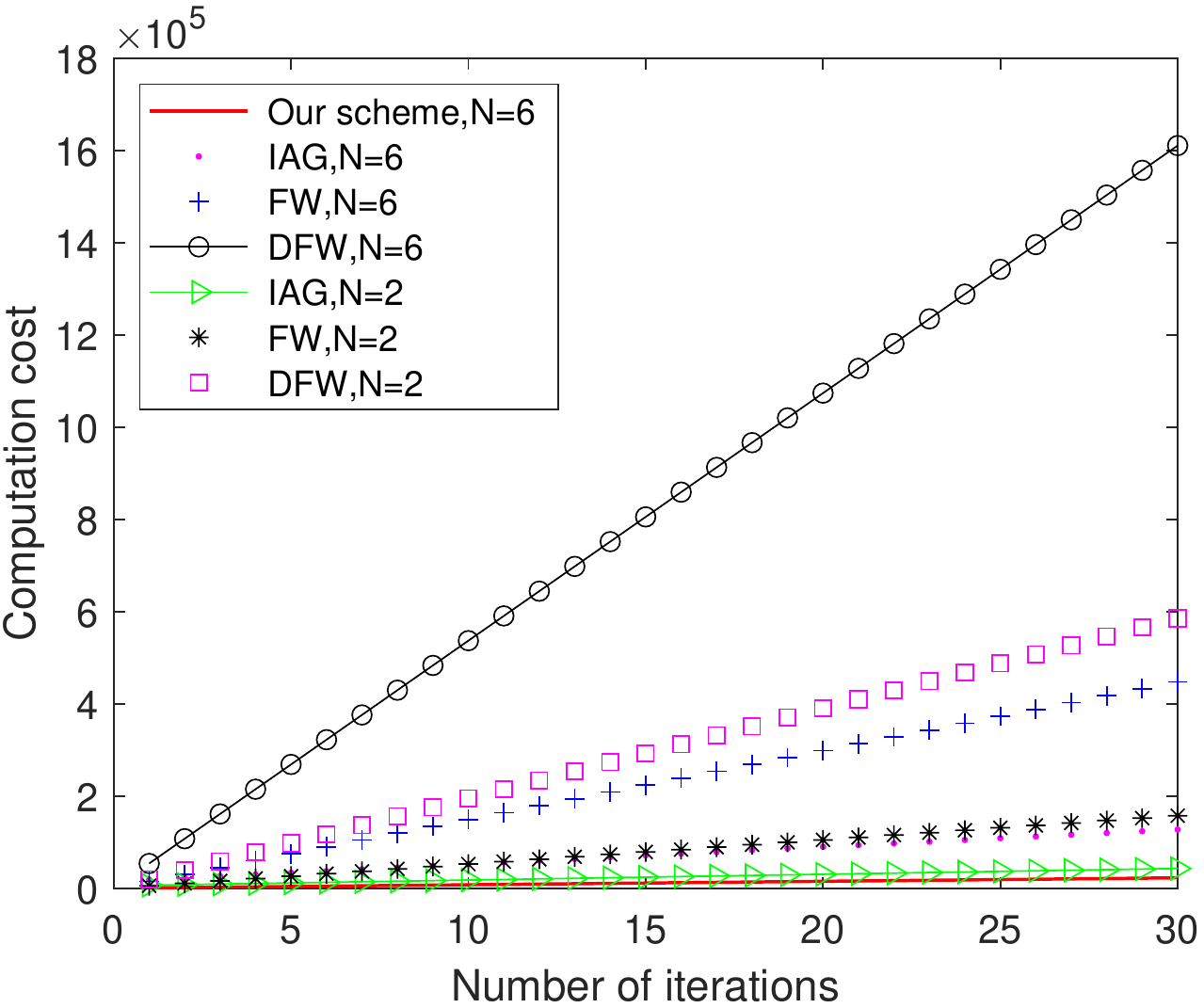}
\caption{Comparison results of the proposed distributed learning-based scheme and the centralized scheme of \cite{25}.}
\end{figure}

Fig. 12 characterizes the performance of the proposed scheme compared to the centralized schemes of \cite{25}, including an incremental aggregated gradient (IAG)-based scheme, Frank-Wolfe (FW)-based scheme, and distributed Frank-Wolfe (dFW)-based scheme. In \cite{25}, the authors proposed a centralized physical layer authentication framework relying on  the help of 6 landmarks to collect the channel estimates of the user to be authenticated. We can observe from Fig. 12 that the computation  cost of our scheme is much lower than that of all schemes of \cite{25}, because our scheme relies on the training models and collaborative processes at the network edge. The cooperative peers do not have to transmit the collected information and training models to the service provider for every iteration. More importantly, the problem is modeled as a convex problem, which is convenient  to solve.
Extra 6 landmarks are used for security purpose in \cite{25}, while the cooperative peers in this paper could be the existing edge devices and edge servers/access points of the network. Hence, our scheme can be widely used in 5G/6G networks.

\section{CONCLUSION}
This paper proposed  a new  collaborative authentication scheme, which relies on multiple edge devices   by distributively sharing authentication information among them. The edge nodes help the service provider to authenticate its users based on local  information collection and processing. Hence,  the authentication accuracy can be improved by relying on  security provision at the network edge, by harnessing multiple cooperating devices, and by utilizing multiple authentication features. A situation-aware secure group update algorithm was developed for updating the cooperative group of  edge nodes and their associated  authentication features autonomously.  To localize the identity spoofer once it is authenticated, an attacker localization algorithm was  proposed.  Our results characterized  the proposed scheme in both indoor and outdoor communication scenarios.  The results demonstrated that the proposed scheme outperforms  the existing centralized authentication schemes.

In our future work, we will focus on further  improving the authentication accuracy of the proposed collaborative authentication scheme in the noisy communication systems operating at low SNRs. We will also explore the employment of more  features for collaborative authentication, which are robust in  dynamic communication environments. The  benefits of edge intelligence will be studied for autonomous  security provision in our future work, which will move the security framework from the center to the network edge, so that the response time and network load of the security provision will be reduced.

\section{APPENDIX}
\subsection{Appendix 1: Proof of Theorem 1}
Given $\mu_{a}$ and $\sigma_{a}^{2}$, the MD rate of the proposed collaborative authentication scheme can be rewritten   according to (5) and (22), which is formulated as
\begin{eqnarray}\nonumber
F_{\rm{MD}}= \int_{0}^{\nu}\frac{1}{\sigma_{a}x\sqrt{2\pi}}e^{-\frac{(\ln x-\mu_{a})^{2}}{2\sigma_{a}^{2}}}dx\\
=\frac{1}{2}(1+erf(\frac{(\ln \nu-\mu_{a})}{\sqrt{2}\sigma_{a}})).~~
\end{eqnarray}
Since there is a trade-off between the MD rate and FA rate of the system, the minimum of $F_{\rm{FA}}$ can be achieved when $F_{\rm{MD}}=\epsilon$.
Hence, the threshold determination process can be formulated   as
\begin{eqnarray}
\frac{1}{2}(1+erf(\frac{(\ln \nu-\mu_{a})}{\sqrt{2}\sigma_{a}}))=\epsilon.
\end{eqnarray}
Then,  $\nu_{\rm{opt}}$ satisfies (25) and  the result of (24) can be obtained.

\subsection{Appendix 2: Proof of Theorem 2}
Given the collaboration cost constraint $\tau$, we have
\begin{eqnarray}
\varphi  \leq\tau
\Rightarrow N \leq [\sqrt{\frac{\tau}{K+1}+\frac{1}{4}}+\frac{1}{2}].
\end{eqnarray}
Since the cooperative peers are trusted, collaborative authentication
can be performed based on the observation of the user's position by
comparing it to the reported position, i.e. to $\bm{a}$.  We need no
less than 3 observations of the location-related features by different
devices to determine the position of a user. Moreover, upon denoting the maximum
number of devices that are available for helping the service provider
at time $t$ by $N_{0}[t]$, the optimal number of cooperative peers is
chosen to be as high as possible to achieve the best authentication
performance. Hence, the result of (27) can be readily derived.

\ifCLASSOPTIONcaptionsoff
  \newpage
\fi

\appendices

~\\

\begin{IEEEbiography}[{\includegraphics[width=1in,height=1.25in,clip,keepaspectratio]{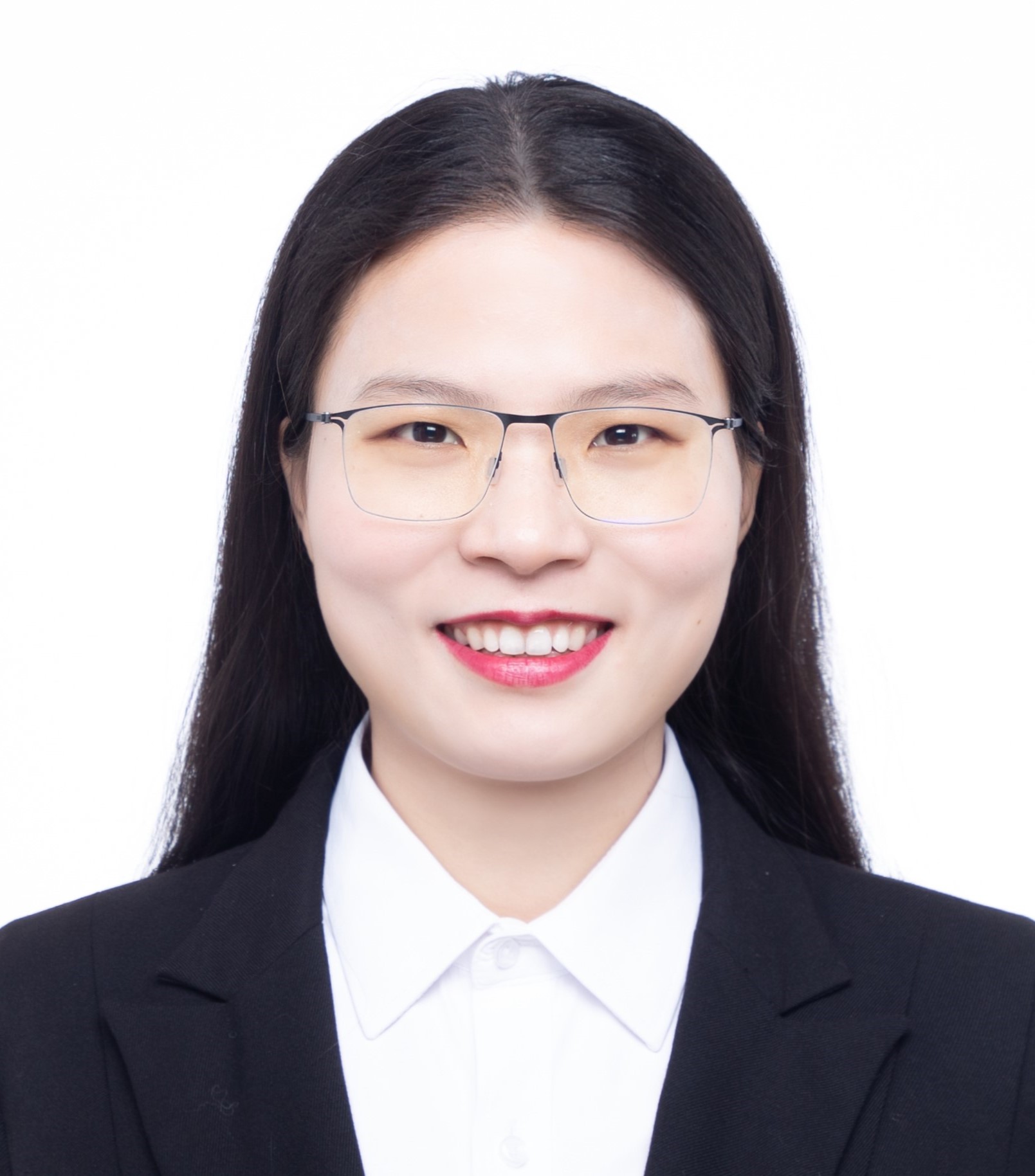}}]{He Fang}
(Member, IEEE) is a full professor
with the School of Electronic and Information Engineering,
Soochow University, China. She received her Ph.D. degree
in Electrical and Computer Engineering from Western University,
Canada, in 2020.
Her research interests include intelligent security
provision, trust management, machine learning, distributed
optimization and collaboration techniques.
She currently serves as a Guest Editor and Topical Advisory Panel Member for several journals, including IEEE Wireless Communications. She  was involved in many IEEE conferences including IEEE GLOBECOM,
VTC,  and ICCC, in different roles such as Session Chair and TPC member. She also served as the Vice-Chair of Communication/Broadcasting Chapter, IEEE London Section, Canada, from Sep. 2019 to Aug. 2021.

\end{IEEEbiography}
\vspace{-3em}
\begin{IEEEbiography}[{\includegraphics[width=1in,height=1.25in,clip,keepaspectratio]{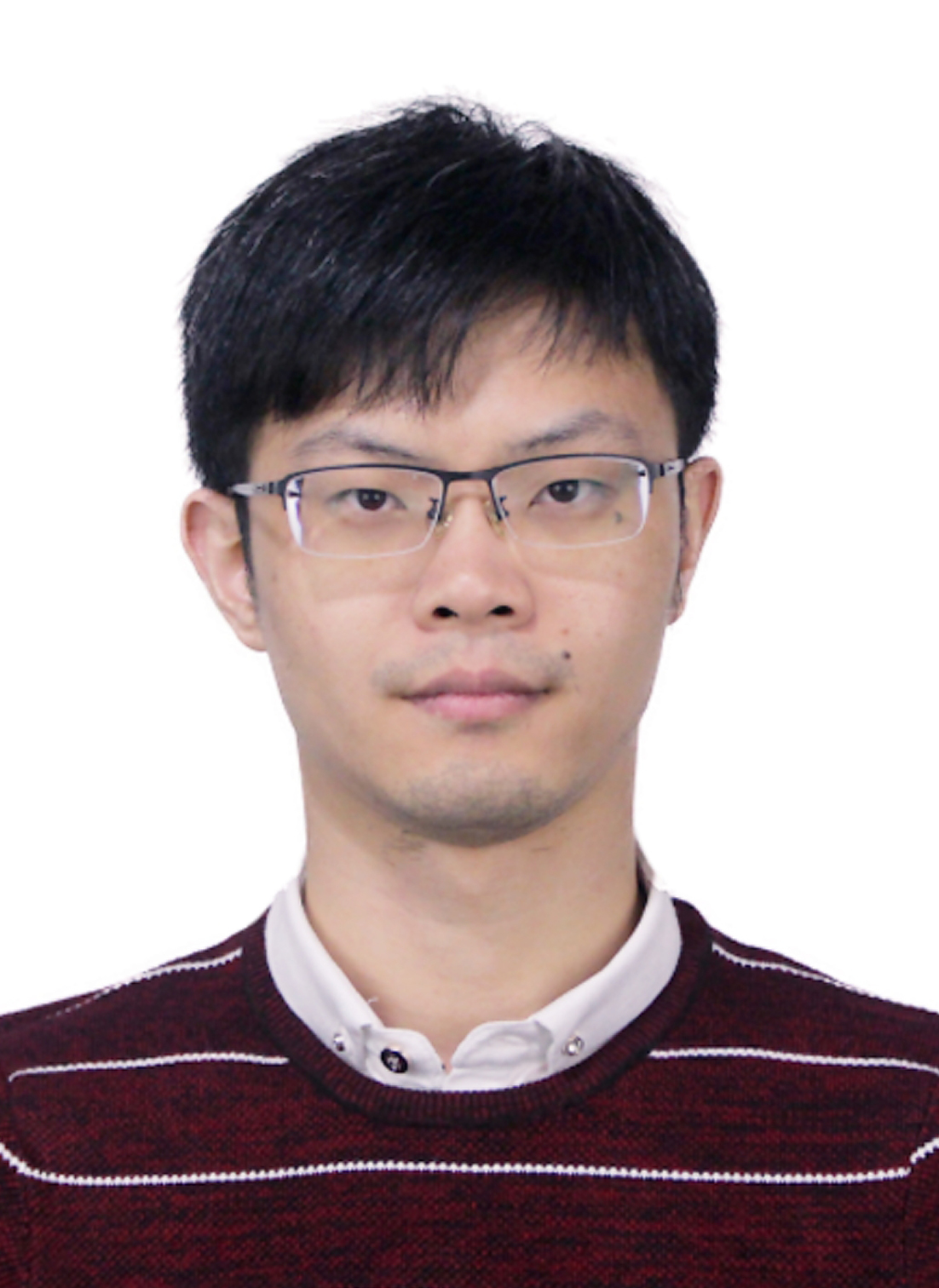}}]{Zhenlong Xiao} (Member, IEEE) received the B.S. degree from the Nanjing University of Posts and  Telecommunications, Nanjing, China, in 2008, the M.S. degree from the Beijing University of Posts and  Telecommunications, Beijing, China, in 2011, and the Ph.D. degree from the Hong Kong Polytechnic University, Hong Kong, in 2015. He is currently an Associate Professor with the Department of Informatics and Communication Engineering, School of Informatics, Xiamen University, Xiamen, China. His research interests include the nonlinear signal processing, graph signal processing, and collaborative signal processing.
\end{IEEEbiography}
\vspace{-3em}
\begin{IEEEbiography}[{\includegraphics[width=1in,height=1.25in,clip,keepaspectratio]{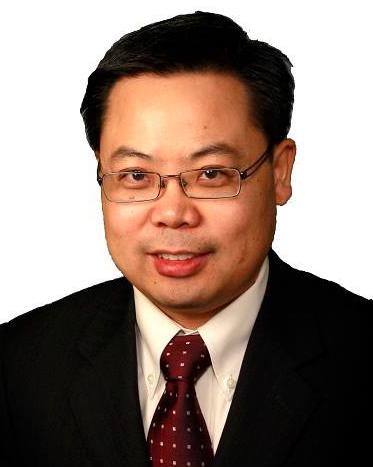}}]{Xianbin Wang} (Fellow, IEEE) is a Professor and Tier-1 Canada Research Chair at Western University, Canada. He received his Ph.D. degree in electrical and computer engineering from the National University of Singapore in 2001.

Prior to joining Western, he was with Communications Research Centre Canada as a Research Scientist/Senior Research Scientist between July 2002 and Dec. 2007. From Jan. 2001 to July 2002, he was a system designer at STMicroelectronics.  His current research interests include 5G/6G technologies, Internet-of-Things, communications security, machine learning and intelligent communications. Dr. Wang has over 500 highly cited journal and conference papers, in addition to 30 granted and pending patents and several standard contributions.

Dr. Wang is a Fellow of Canadian Academy of Engineering, a Fellow of Engineering Institute of Canada, a Fellow of IEEE and an IEEE Distinguished Lecturer. He has received many prestigious awards and recognitions, including IEEE Canada R.A. Fessenden Award, Canada Research Chair, Engineering Research Excellence Award at Western University, Canadian Federal Government Public Service Award, Ontario Early Researcher Award and six IEEE Best Paper Awards. He currently serves/has served as an Editor-in-Chief, Associate Editor-in-Chief, Editor/Associate Editor for over 10 journals. He was involved in many IEEE conferences including GLOBECOM, ICC, VTC, PIMRC, WCNC, CCECE and CWIT, in different roles such as general chair, symposium chair, tutorial instructor, track chair, session chair, TPC co-chair and keynote speaker. He has been nominated as an IEEE Distinguished Lecturer several times during the last ten years. Dr. Wang was the Chair of IEEE ComSoc Signal Processing and Computing for Communications (SPCC) Technical Committee and is currently serving as the Central Area Chair of IEEE Canada.
\end{IEEEbiography}
\vspace{-3em}
\begin{IEEEbiography}[{\includegraphics[width=1in,height=1.25in,clip,keepaspectratio]{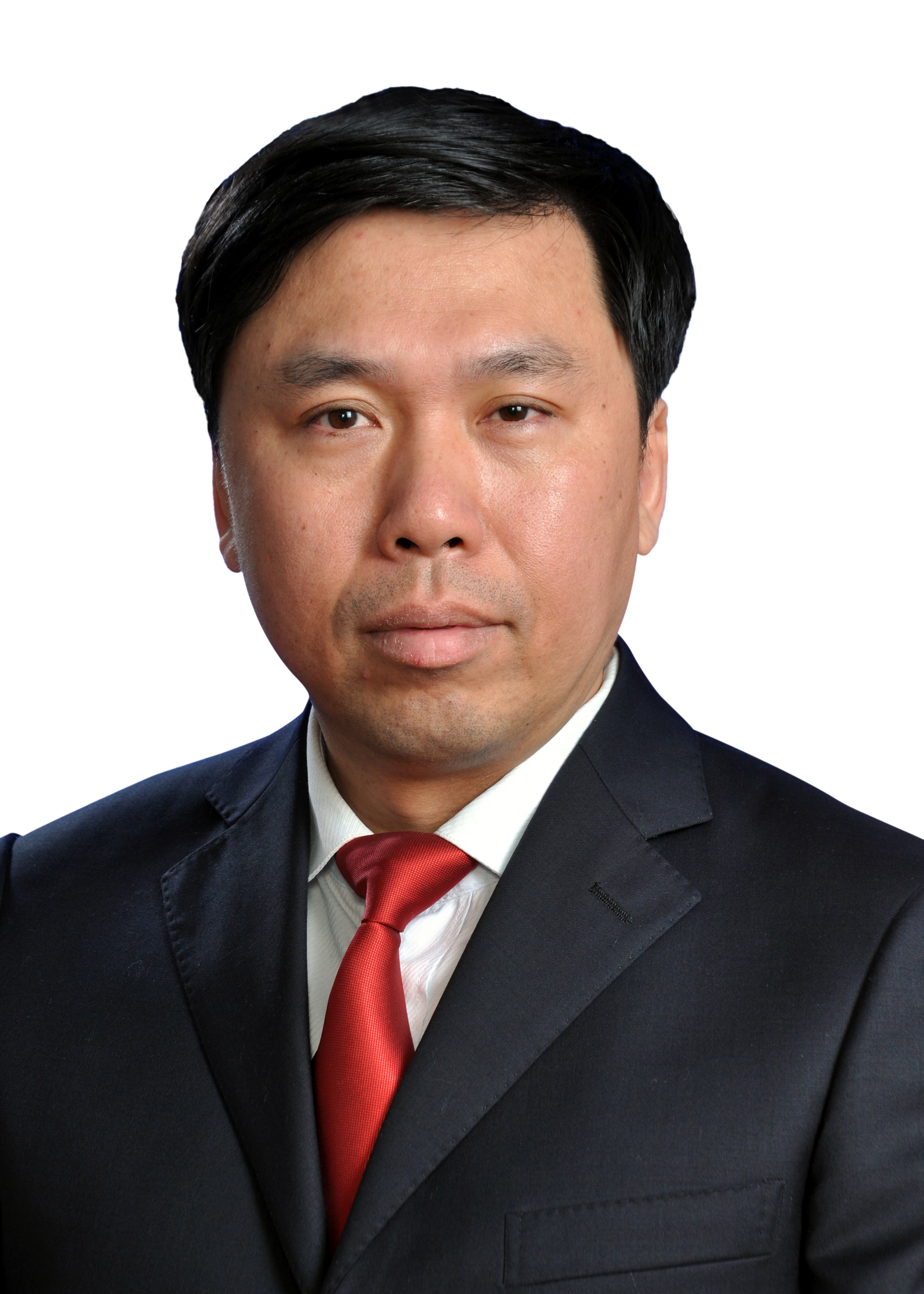}}]{Li Xu} (Member, IEEE) is a Professor with the College of Computer and Cyber Security, Fujian Normal University, Fuzhou. He received the Ph.D. degree from the Nanjing University of Posts and Telecommunications, Nanjing, China, in 2004. He currently is the Dean of the College of Computer and Cyber Security and the Director of Fujian Key Laboratory of Network Security and cryptography. His research interests include network and information security, wireless networks and communication, Big data and intelligent information in complex networks. He has authored or coauthored more than 180 papers in international journals and conferences, including IEEE Transactions on Information Forensics and Security, IEEE Transactions on Computer, IEEE Transactions on Dependable and Secure Computing, IEEE Transactions on Parallel and Distributed Systems.
\end{IEEEbiography}
\vspace{-3em}
\begin{IEEEbiography}[{\includegraphics[width=1in,height=1.25in,clip,keepaspectratio]{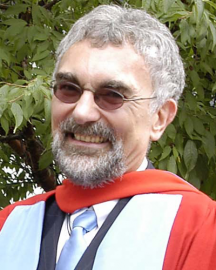}}]{Lajos Hanzo} (FIEEE'04)
 received the  Honorary Doctorates degree from the Technical  University of Budapest and Edinburgh University.   He is a Foreign Member of the Hungarian Science-Academy and a fellow of the Royal Academy
of Engineering (FREng), IET, and EURASIP.  He was a recipient of the IEEE Eric Sumner   Technical Field Award. For more information,   see (http:$//$www-mobile.ecs.soton.ac.uk,
https:$//$en.wikipedia.org$/$wiki$/$Lajos$\_$Hanzo).

\end{IEEEbiography}

\end{document}